\documentclass[preprint]{aastex}

\shorttitle{Galactic field in the line of sight of Tombaugh~1}
\shortauthors{Carraro et al.}

\begin{document}

\title{Galactic structure in the outer disk: the field in the line of sight to the intermediate-age open cluster Tombaugh~1\thanks{Based on observations
carried out at Las Campanas Observatory, Chile (program ID CN009B-042) and Cerro Tololo Inter-American Observatory.}}


\author{Giovanni Carraro\altaffilmark{1,2}}
\affil{Dipartimento di Fisica e Astronomia, Universit\'a di Padova\\
Vicolo Osservatorio 3 \\
I-35122, Padova, Italy}

\and

\author{Joao Victor Sales Silva\altaffilmark{1,3}}
\affil{Observatorio Nacional/MCT\\
Rua Gen. Jos\'e Cristino 77\\
20291-400, Rio de Janeiro, Brazil}

\and

\author{Christian Moni Bidin}
\affil{Instituto de Astronomia, Universidad Catolica del Norte\\
Av. Angamos 0610, Casilla 1280\\
Antofagasta, Chile}

\and

\author{Ruben A. Vazquez}
\affil{Instituto de Astrofisica de La Plata \\
CONICET/ UNLP, Paseo del Bosque s/n\\
La Plata, Argentina}

\altaffiltext{1}{ESO, Alonso de Cordoba 3107, Santiago de Chile, Chile}
\altaffiltext{2}{giovanni.carraro@unipd.it}
\altaffiltext{3}{Instituto de Astronomia, Universidad Catolica del Norte, Antofagasta, Chile}

\begin{abstract}
We employ optical photometry and high-resolution spectroscopy to study a field toward the open cluster Tombaugh~1, where we identify  a complex population mixture, that we describe in terms of young and old Galactic thin disk. Of particular interest is the spatial distribution of the young  population, which consists of dwarfs with spectral type as early as B6, and distribute in a  {\it blue plume} feature in the colour-magnitude diagram.  For the first time we confirm 
spectroscopically that most of these stars are  early type stars, and not blue stragglers nor halo/thick disk sub-dwarfs.
Moreover, they are not evenly distributed along the line of sight, but crowd at heliocentric distances between 6.6 and 8.2 kpc. We compare these results with present-day understanding of the spiral structure of the Ga;axy and suggest that they traces the outer arm.  This range in distances challenges current
Galactic models  adopting a disk cut-off  at 14 kpc from the Galactic center.  The young dwarfs overlap in space with an older component which identifies the old Galactic thin disk.  Both young and old populations are confined in space since the disk is warped at the latitude and longitude of Tombaugh~1. The main effects of the warp are that the line of sight intersects the disk and entirely crosses it at the outer arm distance, and  that  there are no traces of the closer Perseus arm, which would then be either un-important in this sector, or located much closer to the formal Galactic plane.
We finally analysed a group of giant stars, which turn out to be located at very different distances, and to possess very different chemical properties, with no obvious relation with the other populations.
\end{abstract}


\section{Introduction} \label{sec:intro}
Colour-magnitude diagrams (CMD) of stellar fields in the third quadrant of the Milky Way have repeatedly
unravel the remarkable complexity of  the stellar populations in the outer Galactic disk.
Beyond any reasonable doubt, two dominating features appear on top the main sequence of the nearby Galactic field: 
a thick main sequence (MS) with a prominent turn off point, and 
made of intermediate age stars poor in metals, and an almost vertical sequence of young blue stars, popularised as
the {\it blue plume} \citep{moi06,carr05,car16}.
This complexity was originally interpreted  as the result of the accretion of a satellite onto the Milky Way disk
\citep{mar04,bel04}, and different age and metallicity sequences described
as different episodes of star formation in an individual stellar system, the Canis Major dwarf galaxy.\\
The {\it blue plume} deserves particular attention, since up to date no general consensus exists about its nature. 
This is because in the vast majority of the cases, only photometric data are available, which are not univocal enough to derive solid estimate of the stars' gravity and temperature, and, in turn, to infer their spectral type and distance.
Because of the partial leverage of photometric data, various interpretations of the {\it blue plumes} are available.
They can be made of genuine blue young stars, and describe either the last episode of star formation in Canis Mayor
\citep{bel04}, or the structure of the outer Galactic young disk, organised in spiral arms \citep{carr05,moi06,vazquez08}. 
Alternatively, these stars can be the blue straggler population of the old, metal poor population we mentioned before. Finally,
they can be mostly hot sub-dwarfs of type O and B, which are known to be present in the general Galactic field, both in the disk and in the halo \citep{car15}.
Obviously, a better scrutiny of these different scenarios can be obtained only via a dedicated spectroscopic study of 
these blue stars.
This is one of the main scope of this work.  We focus here on the line of sight toward the loose open cluster Tombaugh~1.
To set the scene, we show in Fig.~1 an optical CMD  of a 20$\times$20 arcmin field, in the V/V-I plane, derived from a novel set of UBV(RI)$_{KC}$ photometry
obtained for the present study.
In this CMD we highlight the features we have been discussing so far with four red boxes. 
Box A encompasses clump stars in the star cluster Tombaugh~1, which we already studied
in \citet{sal16}, and are not relevant for this study.  We remind the reader that Tombaugh~1  turned out to be an intermediate age ($\sim$ 1 Gy) open cluster, with  a metallicy of $[Fe/H]=-0.11\pm$0.02, and at 2.6 kpc from the Sun.
Boxes B, C, and D are on the other hand central for the present investigation. Box C encompasses
a group of blue stars part of the {\it blue plume} feature. Box  D includes stars belonging to a thick blue MS whose turn off
point (TO) is located at $V\sim$ 19, $(V-I)\sim$0.9. Finally, box B is composed by a clump of scattered stars, possibly red giant stars belonging to the
same stellar population of Box D.
The aim of this work is to characterise these three different groups, and establish any possible relation among them.
We selected a sample of stars inside each of these boxes and
obtain for them high resolution spectroscopy, which we are going to present and analyse in tandem with the broad band optical photometry.\\

\noindent
The paper is organised as follows: In Sect.~2 we present the observational material, both photometric and spectroscopic.
Section~3 is devoted to the analysis of the various photometry diagrams, and the derivation of blue plume star individual reddening and distance.
A detailed discussion of the spectroscopic data is performed in Section.~4, and in
Section~5 we focus on the results of the abundance analysis of box B stars. The discussion of our results, and the conclusions of this work are provided in Section~6.

\section{Observations and data reduction}

\subsection{Photometry}
The region of interest has been observed with the Y4KCAM camera
attached to the 1.0m telescope, which is operated by the SMARTS consortium\footnote{{\tt http://http://www.astro.yale.edu/smarts}}
and located  at Cerro Tololo Inter-American Observatory (CTIO). This camera
is equipped with an STA 4064$\times$4064 CCD with 15-$\mu$ pixels, yielding a scale of
0.289$^{\prime\prime}$/pixel and a field-of-view (FOV) of $20^{\prime} \times 20^{\prime}$ at the
Cassegrain focus of the CTIO 1.0m telescope. The CCD was operated without binning, at a nominal
gain of 1.44 e$^-$/ADU, implying a readout noise of 7~e$^-$ per quadrant (this detector is read
by means of four different amplifiers)\footnote{QE and other detector characteristics can be found at:
http://www.astronomy.ohio-state.edu/Y4KCam/detector.html}. As an illustration 
we show a V-band frame in Fig.~2.\\

\noindent
The observational data were acquired on the night of January 30, 2008, as summarised in Table~\ref{table1}. 
We observed Landolt's SA~98 $UBV(RI)_{KC}$ standard stars area \citep[][see Table~1]{lan92}, to tie our $UBVRI$
instrumental system to the standard system. The average seeing was 1.0$\arcsec$.\\

\begin{table}
\tabcolsep 0.05truecm
\caption{Log of $UBVRI$ photometric observations.}
\label{table1}
\begin{tabular}{lcccc}
\hline
\noalign{\smallskip}
Target& Date & Filter & Exposure (sec) & airmass\\
\noalign{\smallskip}
\hline
\noalign{\smallskip}
Tombaugh~1 & 30 January 2008 & U & 10,20,100,200,600,1500    &1.28$-$1.52\\
           &                 & B & 5,10,100,200,1600,1500    &1.15$-$1.20\\
           &                 & V & 5,10,60,120,600,1200      &1.01$-$1.21\\
           &                 & R & 3x5,10,60,120,600,1200    &1.02$-$1.15\\
           &                 & I & 5,4x10,100,200,600,1200   &1.03$-$1.24\\
SA~98      & 30 January 2008 & U & 2x10,200,2x400            &1.15$-$2.21\\
           &                 & B & 10,100,2x200              &1.15$-$2.36\\
           &                 & V & 10,50,2x100               &1.16$-$2.53\\
           &                 & R & 10,50,2x100               &1.17$-$2.61\\
           &                 & I & 10,50,2x100               &1.16$-$2.46\\
\noalign{\smallskip}
\hline
\end{tabular}
\end{table}

Our $UBVRI$ instrumental photometric system was defined by the use of a standard broad-band
Kitt Peak $BVR_{kc}I_{kc}$ set in combination with a U+CuSO4 $U$-band filter\footnote{Transmission
curves for these filters can be found at http://www.astronomy.ohio-state.edu/Y4KCam/filters.html}.
To determine the transformation from our instrumental system to the standard Johnson-Kron-Cousins
system, we observed 46 stars in area SA~98 \citep{lan92} multiple times, and with different
air-masses ranging from $\sim$1.2 to $\sim$2.3.  Field SA~98 is very advantageous, as it
includes a large number of well observed standard stars, and it is completely covered by
the CCD's FOV.  Furthermore, the standard's color coverage is very good, being:
$-0.5 \leq (U-B) \leq 2.2$; $-0.2 \leq (B-V) \leq 2.2$ and $-0.1 \leq (V-I) \leq 6.0$.\\

  \begin{figure}
   \centering
   \label{cmd_tom1}
   \includegraphics[width=\columnwidth]{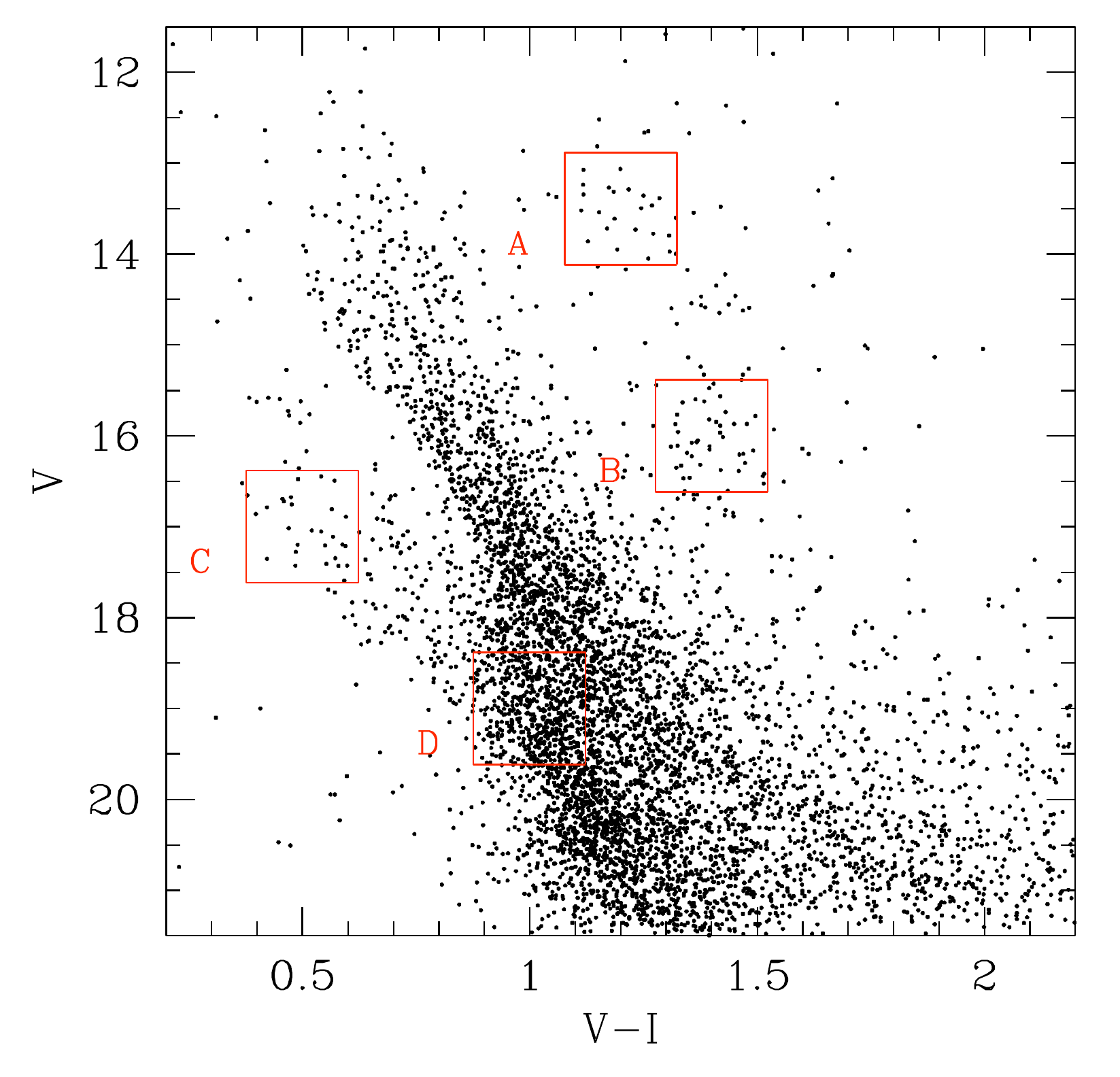}
   \caption{A color-magnitude diagram of  the region under study, with highlighted the areas where spectroscopy has been conducted.} 
    \end{figure}
    
   \begin{figure}
   \centering
   \label{To1}
   \includegraphics[width=\columnwidth]{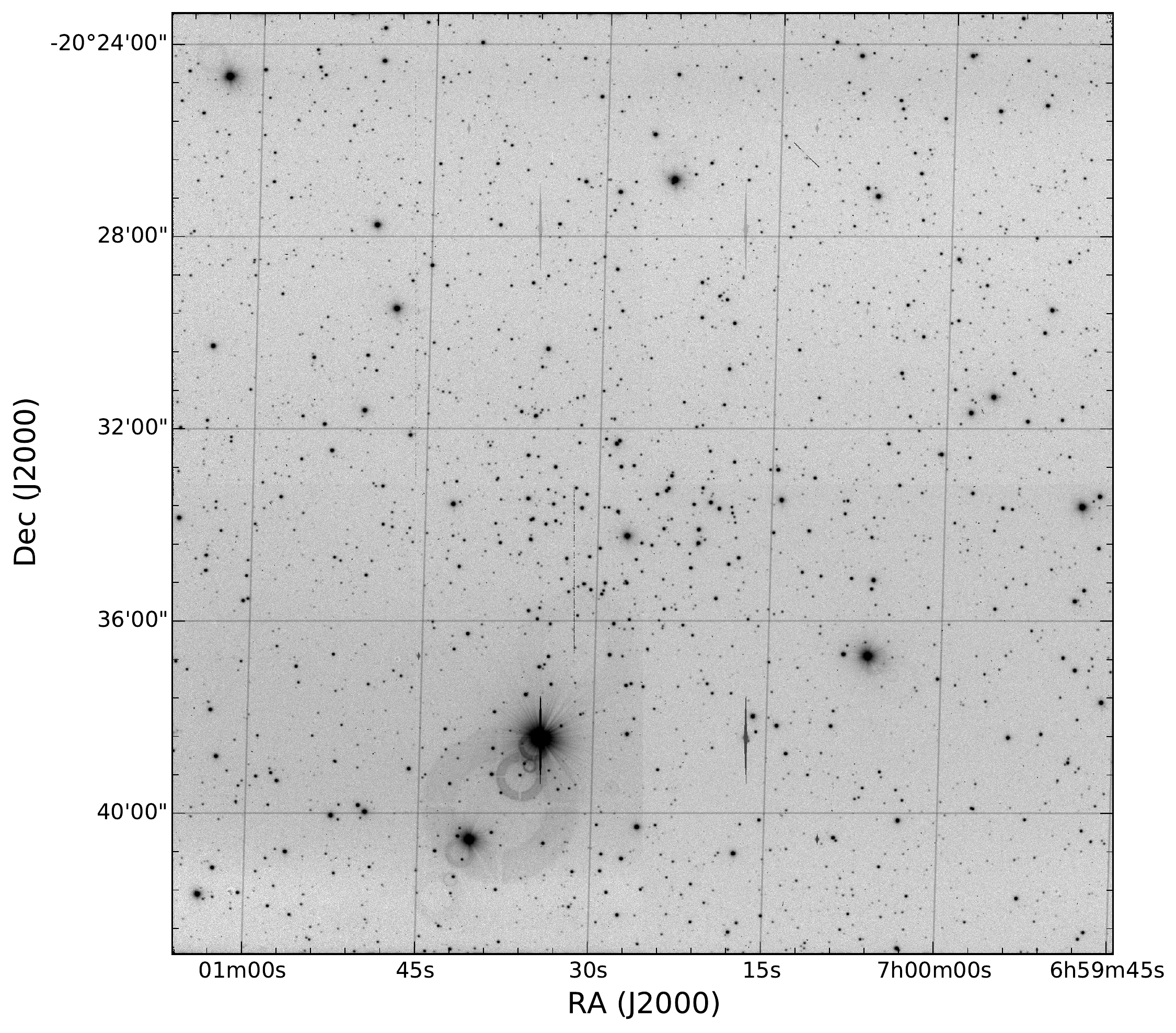}
   \caption{A  900 secs V band frame of the area covered by this study.} 
    \end{figure}
       
\subsubsection{Data Reduction}

Basic calibration of the CCD frames was done using the IRAF\footnote{IRAF is distributed
by the National Optical Astronomy Observatory, which is operated by the Association
of Universities for Research in Astronomy, Inc., under cooperative agreement with
the National Science Foundation.} package CCDRED. For this purpose, zero-exposure
frames and twilight sky flats were acquired every night.  Photometry was then performed
using the IRAF DAOPHOT and PHOTCAL packages. Instrumental magnitudes were extracted
following the point spread function (PSF) method (Stetson 1987). A quadratic, spatially
variable, master PSF (PENNY function) was adopted. The PSF photometry was finally
aperture-corrected, filter by filter. Aperture corrections were determined by performing
aperture photometry for a suitable number (typically 20 to 40) of bright stars selected across the whole field.
These corrections were found to vary between 0.105 and 0.315 mag, depending on the filter.\\

\subsubsection{Final photometry}

Our final photometric catalogs consist of 3275 entries with $UBV(RI)_{KC}$ measurements down to 
$V \sim $ 22 for Tombaugh~1  Many more entries are available when we include star
not having $U$ magnitude.

After removing both saturated stars and stars having only a few measurements in 
the catalog of \citet{lan92},
our photometric solutions for a grand total of 183 measurements in $U$ and $B$, and of 206
measurements in $V$, $R$ and $I$, are given by:\\

\noindent
$ U = u + (3.279\pm0.010) + (0.47\pm0.01) \times X - (0.030\pm0.016) \times (U-B)$ \\
$ B = b + (2.033\pm0.012) + (0.29\pm0.01) \times X - (0.110\pm0.012) \times (B-V)$ \\
$ V = v + (1.673\pm0.007) + (0.16\pm0.01) \times X + (0.022\pm0.007) \times (B-V)$ \\
$ R = r + (2.768\pm0.007) + (0.10\pm0.01) \times X + (0.053\pm0.007) \times (V-R)$ \\
$ I = i + (2.674\pm0.011) + (0.08\pm0.01) \times X + (0.048\pm0.008) \times (V-I)$ \\

\noindent
The final {\it r.m.s} of the fitting was 0.073, 0.069, 0.035, 0.030, and 0.030 in $U$, $B$, $V$, $R$ and $I$.\\

\noindent
Global photometric errors were estimated using the scheme developed by \citet[][Appendix~A1]{pat01},
which takes into account the errors in the PSF fitting
procedure (i.e. from ALLSTAR), and the calibration errors (corresponding to the zero point,
color terms, and extinction errors). In Fig.~3 we present the global photometric error trends 
plotted as a function of $V$ magnitude. Quick inspection indicates that stars brighter than
$V \approx 21$ mag have errors much lower than 0.10~mag both in color and in  magnitude, apart
from the $(U-B)$ color.\\

   \begin{figure}
   \centering
   \label{Fig3}
   \includegraphics[width=\columnwidth]{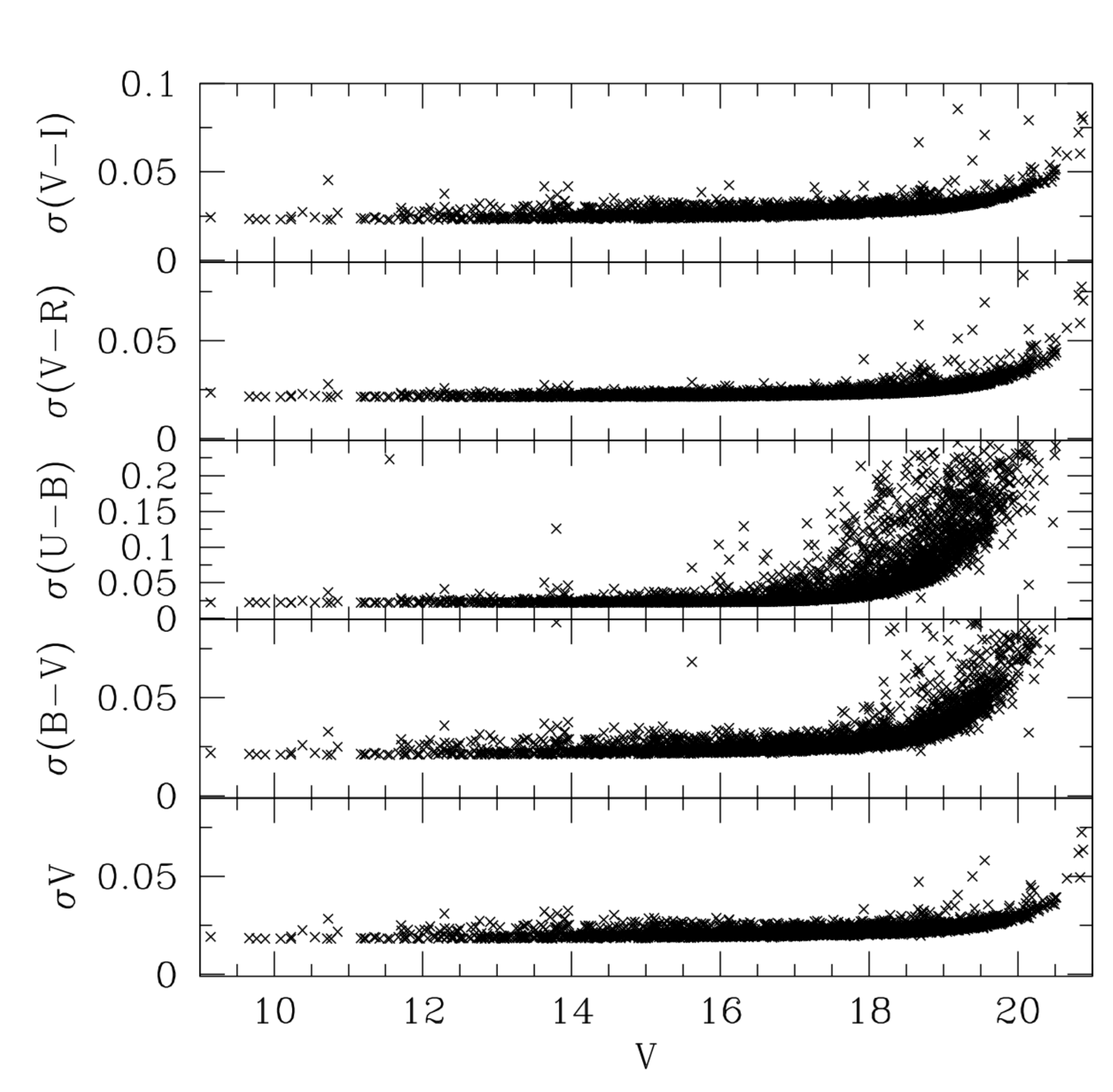}
   \caption{Trends of global photometric errors in color and magnitude as a function of the $V$ mag.} 
    \end{figure}

\subsection{Astrometry}

For approximately three-hundred stars in our photometric catalog J2000.0 equatorial coordinates
are available from the Guide Star Catalogue\footnote{Space telescope Science Institute, 2001, The Guide
Star Catalogue Version 2.2.02.}, version 2 (GSC-2.2, 2001). Using the SkyCat tool at ESO and
the IRAF tasks {\it ccxymatch} and {\it ccmap}, we first established a transformation between our $(X,Y)$
pixel coordinates (from ALLSTAR) and the International Celestial Reference Frame.
These transformations  have an ${\it r.m.s.}$ value of typically 0.15$^{\prime\prime}$.
Finally, using the IRAF task {\it cctran}, we computed J2000.0 coordinates for all objects in our
catalog.

\subsection{Spectroscopy}

During the nights of 2010 January 5, 6, 9 and 10, we observed 40~stars of the field towards the open cluster Tombaugh~1 (10 stars
from box A and B, 11 stars from box C and 9 stars from box D, see Fig.~1 on Cerro Manqui at the Las Campanas Observatory using
the {\it Inamori-Magellan Areal Camera \& Spectrograph} \citep[IMACS,][]{dre06}, attached to the 6.5m Magellan Telescope. The ten
potential cluster stars of box~A were studied in \citet{sal16}. For the stars of boxes A, B and C, we used the multi-object
echelle (MOE) spectroscopic mode, while the spectra of the box D stars were obtained using the multi-object mode with the grating
600 lines/mm (G600-8.6). The spectra have a resolution of R$\approx$20000 and R$\approx$5260 in case of  the MOE and G600 mode,
respectively. In both spectroscopic modes the wavelength ranges of stellar spectra vary according to the position of the star in
the observation mask, but usually it goes from 4200~\AA{} to 9100~\AA{} for the MOE mode, while for the G600 mode the range
is from 3650~\AA{} to 6750~\AA{}. The IMACS detector is a mosaic of eight CCDs with gaps of 0.93mm between them, causing small
gaps in stellar spectra. The exposure times for the stars of the boxes B, C and D were 9000s, 14400s and 6300s, divided in 3, 4 and
2~exposures, respectively.
Table~\ref{t_specdata} gives some information about the observed stars: identification (IDs), equatorial coordinates, $V$ and $(V-I)$ from our photometry, and
spectral signal-to-noise (S/N) at 6000 \AA{}. The identification system for all stars analysed in this work refers to identification of
stars in our photometry. The nominal S/N ratio was evaluated by measuring with IRAF the {\it rms} flux fluctuation in selected continuum
windows.\\

\noindent
The reduction of the spectra was performed in a standard manner under IRAF as described in details in \citet{sal16}. The eight CCDs were de-biased and
flat-fielded separately with the task {\it ccdproc}, combined in a single frame with {\it imcreate} and {\it imcopy}, then the
spectra were optimum-extracted \citep{hor86} with {\it doecslit} ({\it doslit} for G600 mode), sky-subtracted with {\it background},
and wavelength calibrated with {\it ecidentify} ({\it identify} for the G600 spectra). The cosmic rays were removed with the IRAF
Laplacian edge-detection routine \citep{dok01}.

\section{Photometric diagrams}
\begin{figure*}
   \centering
   \plottwo{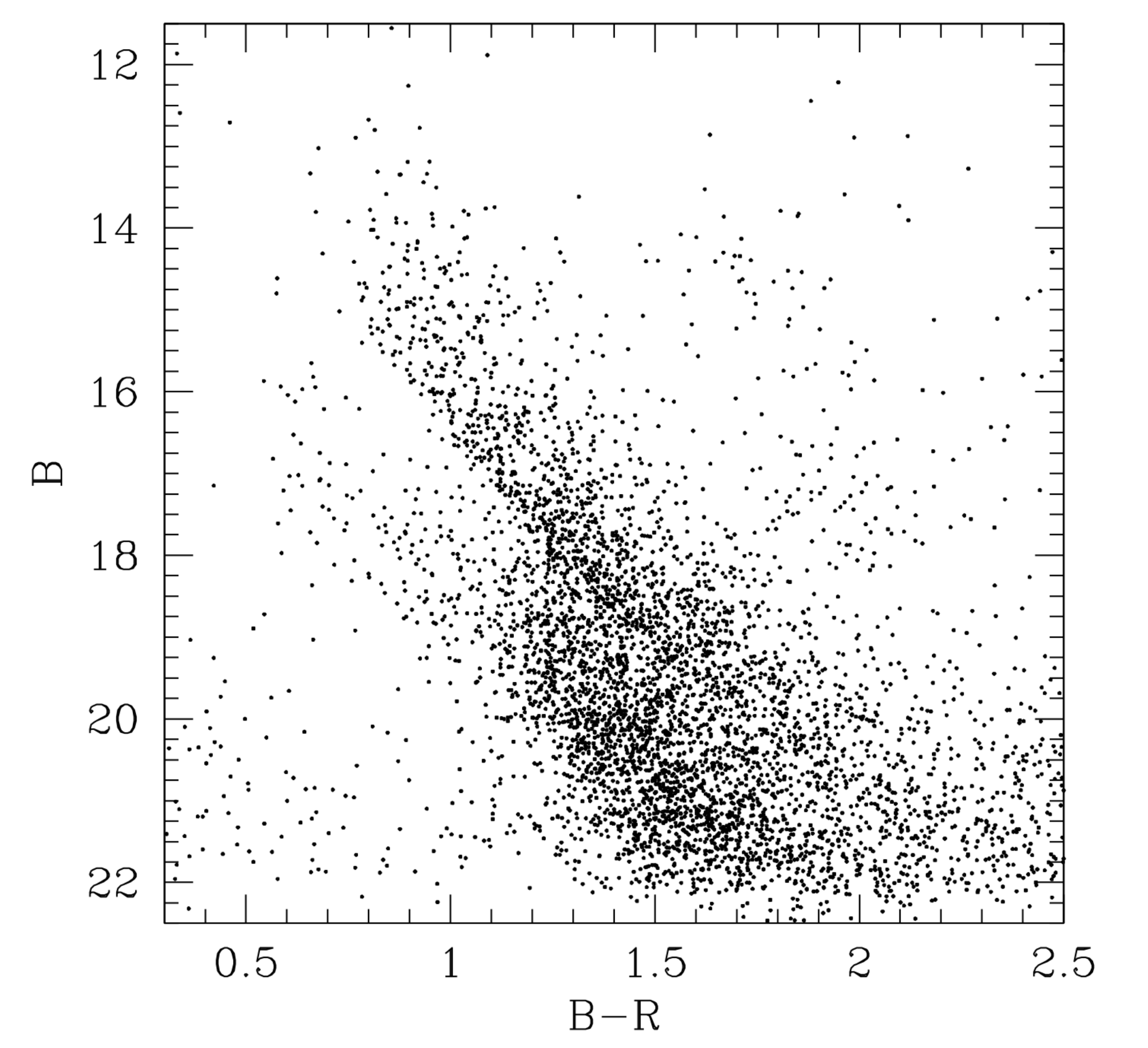}{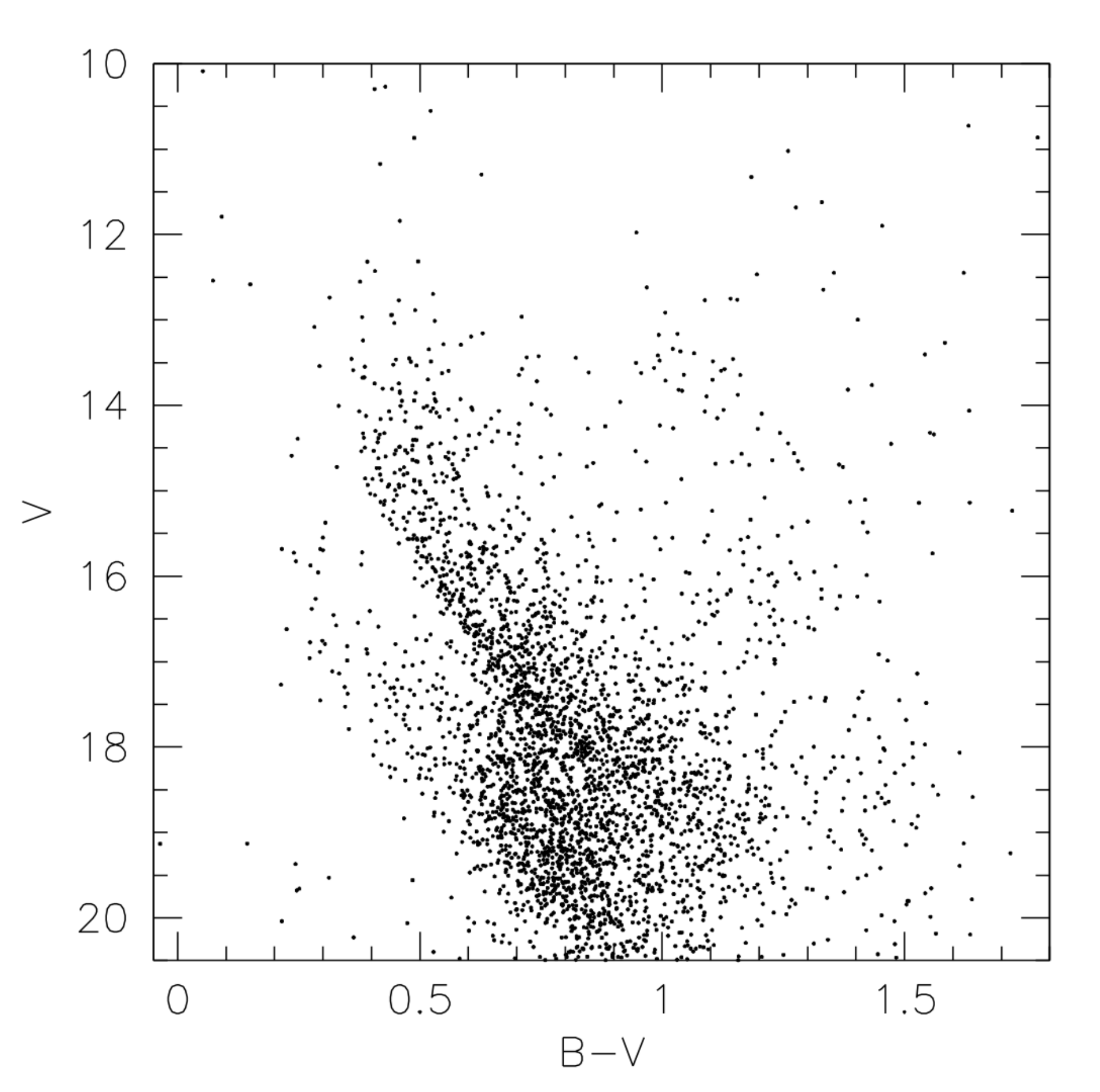}
   \caption{CMDs of the stars in the line of sight to the star cluster Tombaugh~1. In the right panel, the CMD has been constructed using stars having simultaneous measures  in U, B, and V only, to make easier the interpretation of Fig.~5.} 
\label{f_cmd2}
    \end{figure*}

We start deriving some insights on the properties of the stellar population in the line of sight to Tombaugh~1 by inspecting a suite of photometric diagrams.
Inspection of the CMDs of  all the stars in the field of view
in Fig.~\ref{f_cmd2} reveals three prominent features: (1) a cluster MS with a TO at $B\sim14.5$ ($V\sim 18.5$) and a handful of scattered red clump stars at $B\sim14.3$  ($V \sim 14.2$)and 
(B-R) $\sim$ 1.2,
which we discussed in \citet{sal16};
(2) a second, thick, well-populated MS with a TO at $B\sim19.5$ that looks like the MS of an intermediate-age/old stellar population;
and (3) a scattered plume of blue stars, in the magnitude range 16--19 that resembles a young stellar population. 
The last feature is very similar to the {\it blue plumes} found in the directions of other clusters \citep[][and references therein]{car16} in the third Galactic quadrant.

  \begin{figure*}
   \centering
   \includegraphics[scale=0.8]{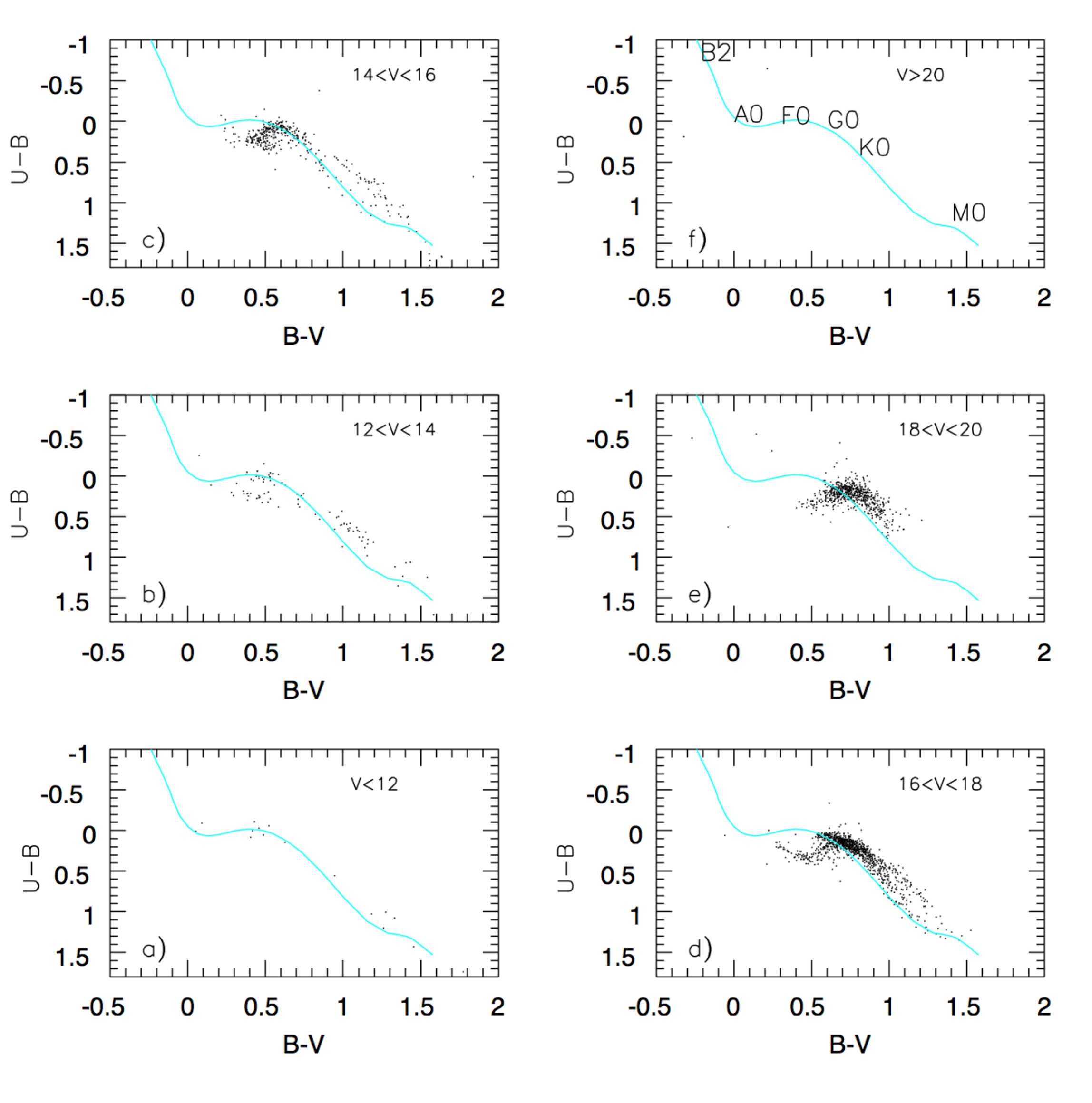}
   \caption{The TCDs of the stars in the line of sight to the star cluster Tombaugh~1 extracted from the CMD (right panel of Fig.~4), by binning in magnitude. Only stars having U, B, and V measures simultaneously are plotted.}
\label{f_tcd1}
    \end{figure*}

\noindent
As amply discussed in \citet{car16} it is quite straightforward to characterise the {\it blue plume}, since it would be composed of supposedly young stars for which a robust handling of their properties is possible withUBV photometry. We start discussing more in detail the CMD, by plotting stars in different  magnitude bins
in the classical two-colour diagram (TCD) U-B/B-V,  as shown in Fig.~\ref{f_tcd1}. A synoptic view of Fig.~4 and Fig.~5
helps the various CMD components to emerge more clearly, as we already  illustrated 
in the past \citep[see, e.g.,][]{car10}. In the various panels the cyan line is an empirical  zero reddening,  zero-age main sequence (ZAMS) from Turner (1976,1979).
The TCD for V$\leq$12 (lower left panel) only shows a few zero reddening stars of different spectral type (from A0 to F-G and M) located close to the Sun. The middle-left panel
is surely more interesting, since on top the clump of nearby un-reddened  F-G stars, it shows two groups of reddened stars (spectral type A and K-M) clearly belonging to the star cluster Tombaugh~1, the first indicating the cluster TO, and the second the red giant clump.  
The upper-left panel is sampling Tombaugh~1 TO and MS, but also shows a few giant stars, both reddened and un-reddened. 
The next panel (lower-right) is by far the most intriguing. One can readily notice an important sequence of early type, reddened stars, totally absent in the previous TCDs,  which runs from approximately mid B spectral type all the way to K-M. 
One can also notice (at $(B-V)$ $\sim$0.5), $(U-B)$ $\sim$ 0.0) a somewhat detached, truncated, less reddened, sequence . This latter is, again, Tombaugh~1 MS, while
the more reddened sequence samples the {\it blue plume} up to spectral type AO, and later starts to sample the thick blue MS whose TO is at (B-V) $\sim$0.8. This can be appreciated by
the density increase of stars about this color in the TCD. Besides, also the giant sequence is dual, and contain zero reddening nearby giants, and more distant, reddened giants, possibly associated with the blue thick MS we just described.
Finally, the middle-right panel only shows stars in the dominant thick blue MS. The last, top-right, panel is only used to indicate the approximate location of different spectral type stars.\\

\begin{figure}
   \centering
   \includegraphics[width=\columnwidth]{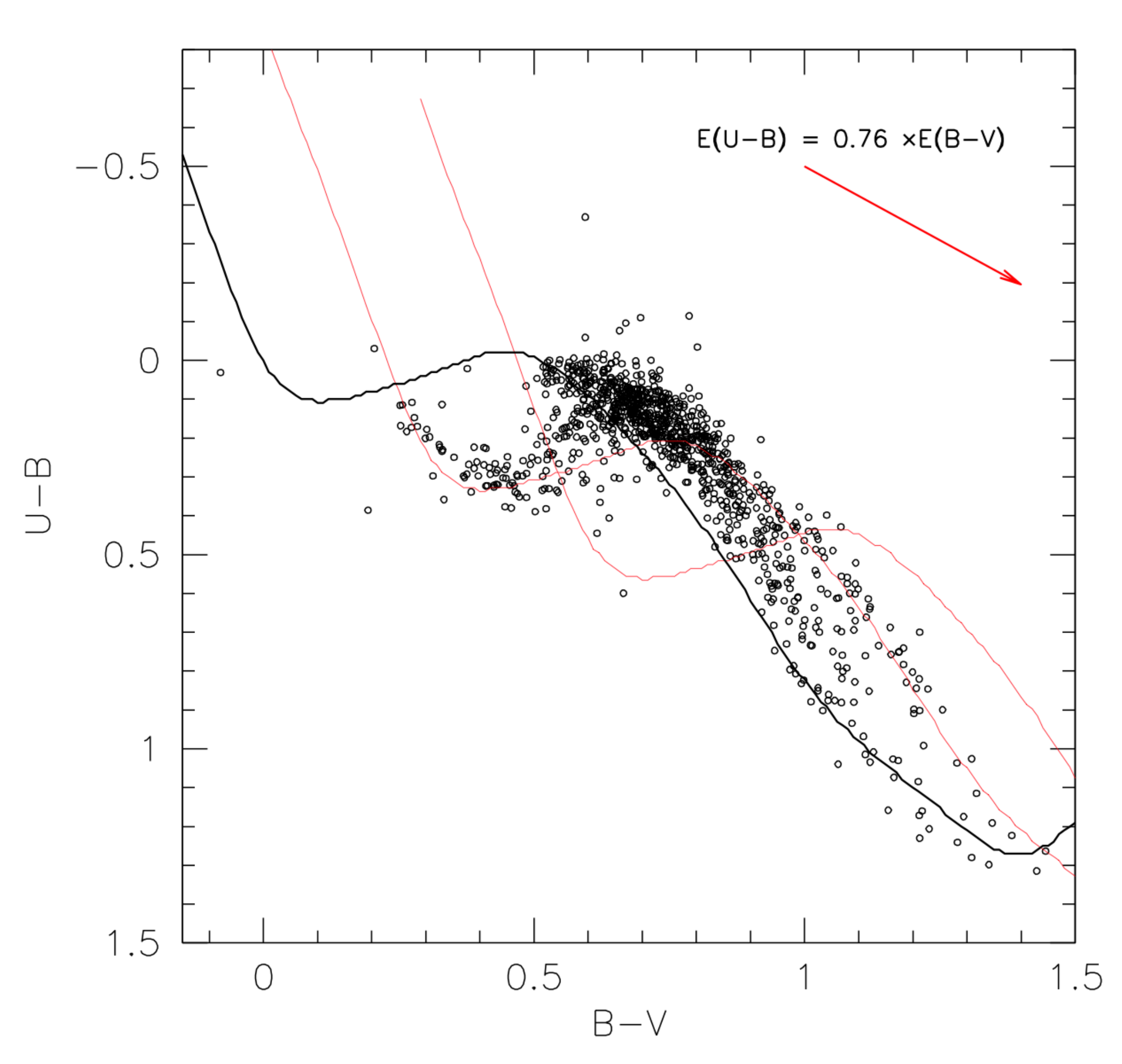}
   \caption{TCDs of the stars in the line of sight to the star cluster Tombaugh~1, and in the magnitude range $16 \leq V \leq 18$. The solid line is a zero reddening, zero age main sequence, while the two red lines encompassing the early type stars are the very same ZAMS, but displaced by 0.4 and 0.6 mag along the reddening vector.} 
\label{f_tcd2}
    \end{figure}

\noindent
We focus now on the lower right panel, and attempt at deriving the properties of the stars which populated the early spectral type branch, and correspond to the {\it blue plume}.
We obtain a reddening solution for these stars using the TCD in the standard way. This is illustrated in Fig.~\ref{f_tcd2}.

The reddening law in the third Galactic quadrant has been recently debated in the literature. As discussed in \citet{car16} and 
earlier by \cite{moitinho01} and \cite{turner76} the reddening law in the third Galactic quadrant seems to deviate from the
normal one, namely it is not described by the standard 
value of 3.1 for $R_V= \frac{A_V}{E(B-V)}$. A value of $R_V$=2.9 would  to be more appropriate for this Galactic sector. 
This is certainly true for Tombaugh 1 line of sight, as discussed by \cite{turner83}. 
Although the level of deviation from the normal law is not large, we prefer to adopt 2.9 in the following.

The solid black line in Fig.~\ref{f_tcd2} is a zero reddening, zero age MS, while the two red lines are the same ZAMS, but shifted by $E(B-V)$ = 0.3 and 0.7 along the reddening line,
which is indicated by the red arrow in the top-right corner of the plot. The two reddened ZAMS  bracket the {\it blue plume} stars, which therefore exhibit a mean reddening $E(B-V)$=0.5$\pm$0.2, indicating a significant amount of variable reddening. We now analyse the reddening distribution of this population by deriving individual stars' reddening.\\

\noindent
To determine reddenings, spectral types and, eventually, distances we then proceed
as follows.
First we derive intrinsic colours using the two relationships:

\begin{equation}
E(U-B) = 0.76 \times E(B-V) ,
\end{equation}

\noindent
and

\begin{equation}
(U-B)_0 = 3.69 \times (B-V)_0 + 0.03.
\end{equation}

\noindent
The intrinsic color (B-V)$_0$ is the positive root of the second order equation one
derives by combining  the above expressions.
Intrinsic colours  ((U-B)$_0$ and (B-V)$_0$) are then directly 
correlated to spectral type, as compiled for instance byTurner (1976,1979). The solution
of the equations above therefore allows us to encounter stars having spectral types
earlier than A0.5. For these stars we know the absolute magnitude M$_{V}$ 
and, from the apparent extinction-corrected magnitude V$_{0}$, we finally infer the photometric distance.\\

\noindent
Starting from the general equation for the distance:

\begin{equation}
(m-M)_o = (m-M)_V - A_V= 5 \times log(Dist) -5 
\end{equation}

\noindent
errors in distances are computed as follows:

\begin{description}
\item $\Delta$ (Dist)  = ln(10) $\times$ Dist $\times$ $\Delta$ [log(Dist)];
\item $\Delta$ [log(Dist)] = $\frac{1}{5}$ $\times$ $\Delta$ V + $\Delta$ (M$_V$) + $\Delta$ (A$_V$)];
\item $\Delta$ (M$_V$) = 0;
\item $\Delta$ (A$_V$) = 2.9 $\times$ $\Delta$ (B-V);
\item $\Delta$ (V) and $\Delta$(B-V)  directly comes from photometry; finally
\item $\Delta$ (Dist) = ln(10) $\times$ Dist $\times$ 1/5 $\times$ [ $\Delta$ V + 2.9 $\times$ $\Delta$ (B-V) ]
\end{description}

 \begin{figure}
   \plottwo{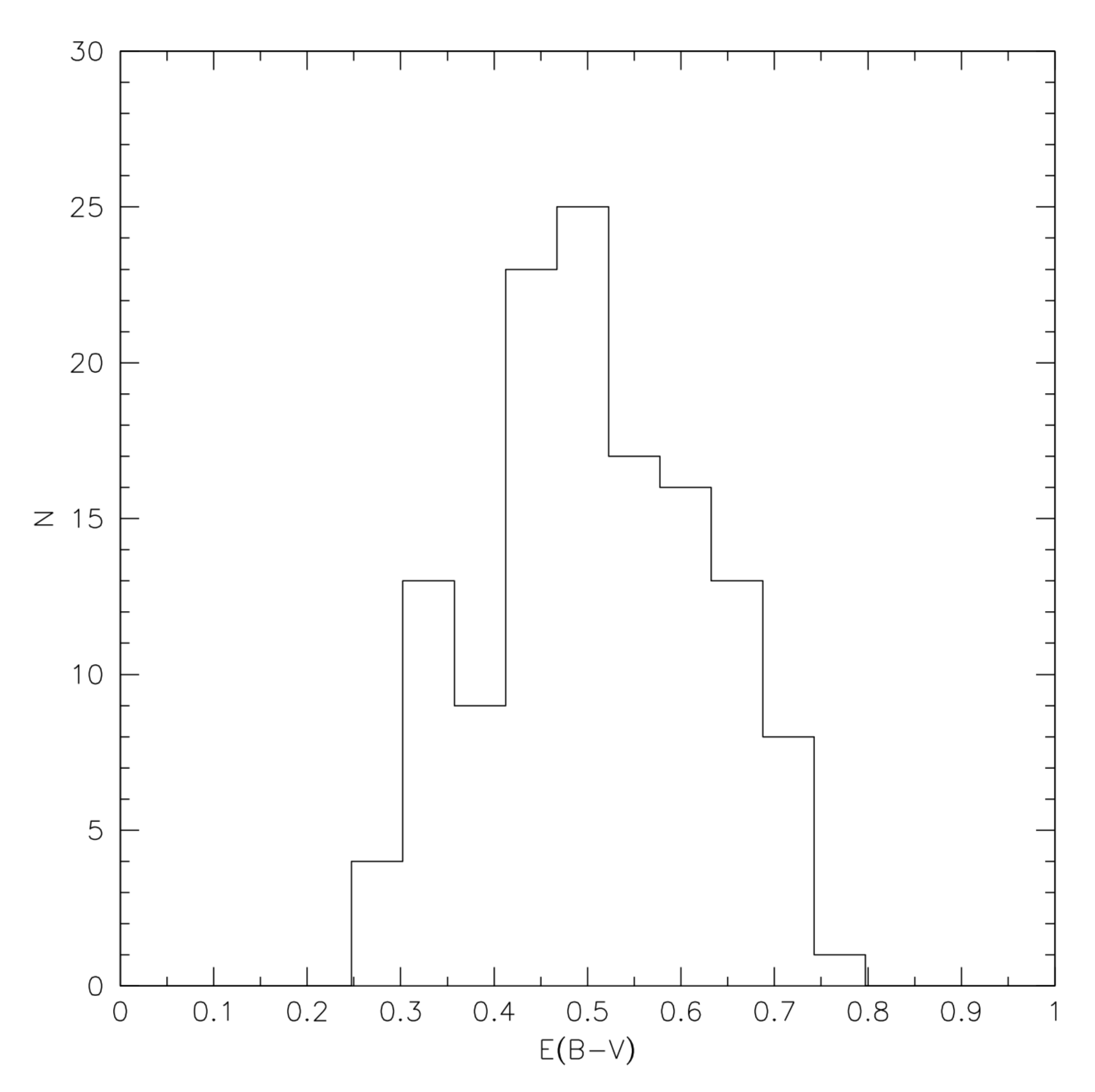}{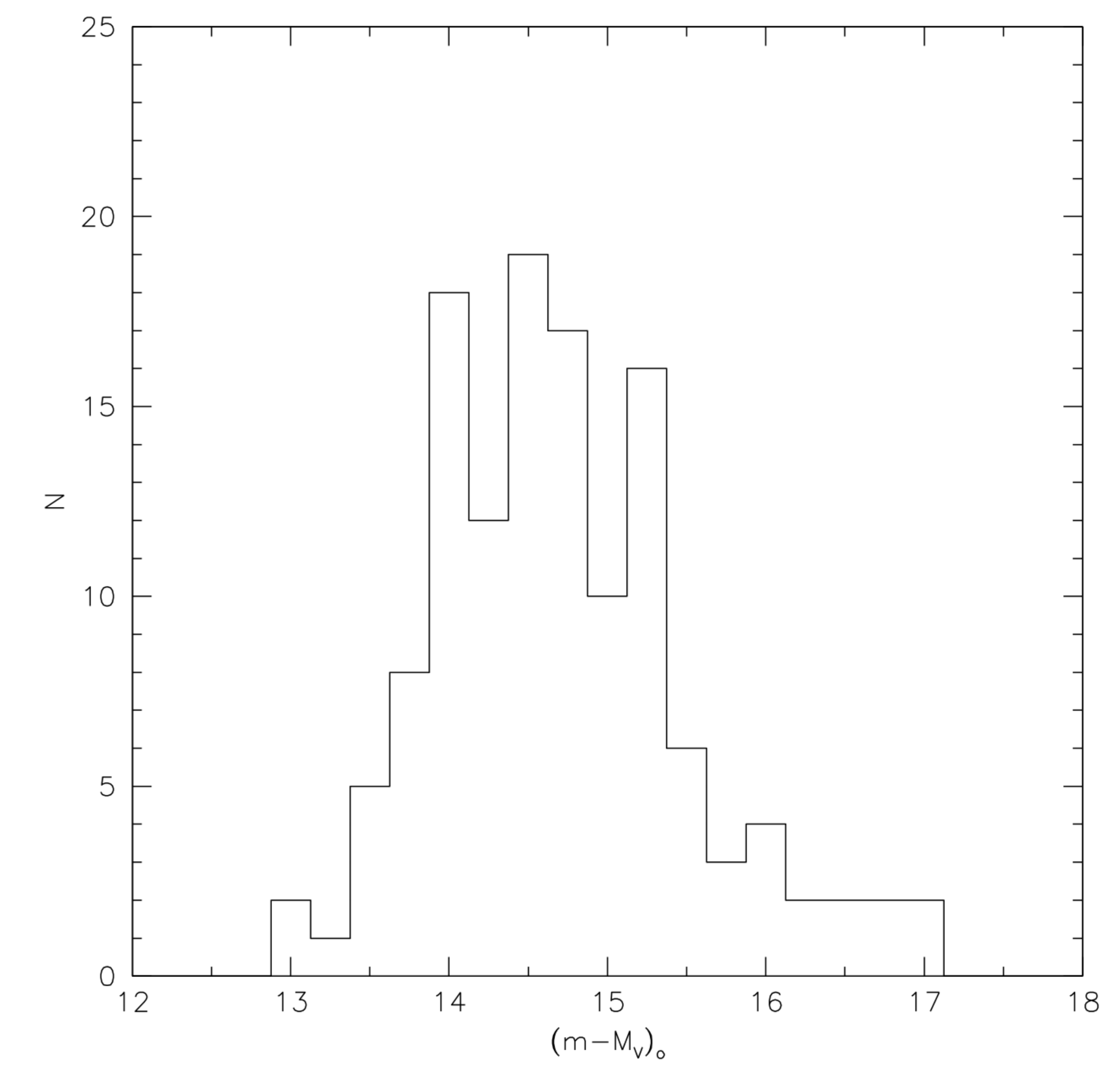}
\label{f_histo}
   \caption{Distribution of individual star reddenings (left panel) and reddening corrected distance moduli (right panel).} 
    \end{figure}

\noindent
The reddening distribution is shown in the left panel of Fig.~7. It is quite broad, and peaks at E(B-V) $\sim$ 0.5.  A gaussian fit yields
the value E(B-V) = 0.55$\pm$0.20.  On the right panel of the same figure we show the distribution of the absolute distance moduli for the
same stars. Most stars are located in the range  14.0 $\leq (m-M)_0 \leq $ 15.5, which implies a distance range 6.3 $\leq d_{\odot} \leq$10.0 kpc.
Errors in distances as computed using the formulae described above, are 0.5 kpc for the closer stars, and up to 1.5 kpc for the most distant stars.
We notice, finally, a group of very distance stars, at $(m-M)_0 \sim 16$, whose distance would be as large as 19$\pm$3.5
kpc. We will return to this group later.

\section{Spectroscopic analysis}

\subsection{Radial velocities}
The radial velocity (RV) of the targets were measured on each single exposure independently. We thus checked for RV variations, and shifted them to laboratory
wavelengths before co-adding the spectra of each star. The zero-point offset of each spectrum was estimated using the {\it fxcor} IRAF task, by cross-correlating
\citep{ton79} the observed telluric band at 6800~\AA{} with that of a FEROS high-resolution solar spectrum collected by us in a previous run \citep{mon12a}.
The heliocentric correction was estimated through the {\it rvcorrect} task, and applied to each measurement. The final RV of each star was obtained from the
weighted mean of the single epochs. Our results are given in Table~\ref{t_atmparamB}.

For box B stars, the line-to-line differences between the observed and laboratory wavelengths of the unblended Fe lines were used to determine the target RV.
The final error was assumed as the largest difference between the three heliocentric radial velocity values multiplied by 0.59
\citep[small sample statistics, see][]{kee62}. The RV of box C stars was estimated by cross-correlating the H$_\alpha$ line with the synthetic template
of a 10\,000~K MS star taken from the \citet{mun05} library. Previous works have shown that the results of cross-correlation are not affected by the exact
choice of the template, and a marginal mismatch between object and template spectral types only enhances the formal uncertainties \citep{mor91,mon11}. When
other strong features were visible in the spectral range, such as the H$_\beta$ line, we verified that they returned identical results within errors. However,
as the availability of these features varied among the targets,  for the sake of consistency we used only the H$_\alpha$ line to derive the
final results. The same procedure was adopted for box D stars, except that the aforementioned FEROS Solar spectrum was used as template.

\begin{table}
\caption{Fundamental information for the spectroscopically observed stars.}
\label{t_specdata}
\begin{tabular}{lccccc}\hline\hline
\multicolumn{6}{c}{Box B}\\\hline
ID     &  RA(2000.0)    &   DEC(2000.0)      &    $V$     &   $(V-I)$  & S/N \\\hline
       &    degree      &    degree          &     mag    &    mag   &     \\\hline
11029  &  105.0619584   &   $-$20.5959712    &   $15.97$  &  $1.446$ & 65  \\
13540  &  105.0867260   &   $-$20.6172278    &   $16.25$  &  $1.393$ & 60  \\
13964  &  105.0912208   &   $-$20.6830230    &   $16.48$  &  $1.457$ & 35  \\
15490  &  105.1050867   &   $-$20.4189783    &   $16.06$  &  $1.324$ & 35  \\
26606  &  105.1789665   &   $-$20.4134974    &   $15.83$  &  $1.428$ & 35  \\
27955  &  105.1880177   &   $-$20.5694871    &   $15.72$  &  $1.381$ & 75  \\
28064  &  105.1887938   &   $-$20.4728788    &   $15.57$  &  $1.392$ & 60  \\
29403  &  105.1989331   &   $-$20.6387084    &   $16.62$  &  $1.512$ & 20  \\
31364  &  105.2138666   &   $-$20.6505511    &   $15.96$  &  $1.475$ & 60  \\
32782  &  105.2261239   &   $-$20.4283761    &   $16.06$  &  $1.416$ & 35  \\
35658  &  105.2524590   &   $-$20.4933638    &   $16.30$  &  $1.458$ & 35  \\\hline\hline
\multicolumn{6}{c}{Box C}\\\hline
 6507  &  105.0154684   &   $-$20.7188840    &   $17.31$  & $0.549$  &  20 \\
 8542  &  105.0351681   &   $-$20.4324223    &   $17.21$  & $0.566$  &  10 \\
 9227  &  105.0426156   &   $-$20.4693299    &   $16.82$  & $0.458$  &  20 \\
12018  &  105.0719883   &   $-$20.4367526    &   $16.46$  & $0.491$  &  15 \\
13279  &  105.0840465   &   $-$20.4816043    &   $16.88$  & $0.676$  &  20 \\
16940  &  105.1177477   &   $-$20.5710991    &   $17.24$  & $0.653$  &  20 \\
24772  &  105.1683352   &   $-$20.4800548    &   $16.77$  & $0.474$  &  15 \\
28816  &  105.1943277   &   $-$20.5676972    &   $16.58$  & $0.489$  &  25 \\
30971  &  105.2109730   &   $-$20.6079883    &   $17.06$  & $0.695$  &  25 \\
31183  &  105.2125614   &   $-$20.5339462    &   $17.30$  & $0.696$  &  15 \\
32089  &  105.2197018   &   $-$20.6357812    &   $16.41$  & $0.659$  &  40 \\\hline\hline
\multicolumn{6}{c}{Box D}\\\hline
 7421  &  105.0244349   &   $-$20.5505733    &   $19.37$  & $1.052$  &  15 \\
 9011  &  105.0398536   &   $-$20.6999266    &   $18.65$  & $0.986$  &  20 \\
 9834  &  105.0492119   &   $-$20.4859526    &   $19.38$  & $1.044$  &  10 \\
11923  &  105.0711463   &   $-$20.4358597    &   $19.10$  & $0.942$  &  10 \\
19341  &  105.1359171   &   $-$20.4658181    &   $19.36$  & $1.045$  &  10 \\
22319  &  105.1533307   &   $-$20.5625042    &   $18.88$  & $1.004$  &  20 \\
23667  &  105.1614867   &   $-$20.4377257    &   $18.46$  & $0.966$  &  15 \\
36132  &  105.2567499   &   $-$20.4139628    &   $18.54$  & $0.986$  &  10 \\
31274  &  105.2133773   &   $-$20.5128144    &   $19.38$  & $1.057$  &  10 \\\hline                                     
\end{tabular}                                       
\end{table}

\subsection{Box B: stellar parameters and chemical abundances}

The lines list used to determine the chemical abundance of Na, Al, Mg, Si, Ca, Ti, Cr, Fe and Ni is the same we used
recently in \citet{sal16}. In Tables~\ref{tabelFea} and \ref{tabellinesa} we show our lines list with excitation potential
($\chi$) and oscillator strength (gf) for all absorption lines analysed in this work. The values of the oscillator
strength adopted for the Fe I and Fe II lines were taken from \citet{lam96} and \citet{cas97}.
The references of the atomic parameters for the other absorption lines are shown in Table~\ref{tabellinesa}.

The chemical abundance of Na, Al, Mg, Si, Ca, Ti, Cr and Ni for the red clump stars (box B) were obtained through the
equivalent widths (EWs) of the absorption lines corresponding to each element. The EWs were measured using the
task {\it splot} in IRAF by fitting the observed absorption line with a Gaussian profile . Absorption lines with EWs
greater than 160 m\AA{} are saturated and were rejected in our analysis due to the impossibility to fit a Gaussian
profile to these lines \citep{per11}. The EWs used to derive the chemical abundance are shown in Tables~\ref{tabelFea} and \ref{tabellinesa}.

The local thermodynamic equilibrium (LTE) model atmospheres of \citet{kur93} and the spectral analysis code MOOG \citep{sne73}
were used to determine the chemical abundances and atmospheric parameters for the stars of box B. The effective
temperature, surface gravity and micro-turbulence velocity were derived using measurements of EWs
for a set of Fe I and Fe II lines shown in Table~\ref{tabelFea}. The unique solution for the effective temperature, surface gravity
and micro-turbulence velocity was obtained simultaneously under the approximations of
excitation and ionisation equilibrium, and independence between the Fe abundance and reduced EW. The
effective temperature was set through the excitation equilibrium determined by zero slope of the trend between the iron
abundance derived from Fe I lines and the excitation potential of the measured lines. The micro-turbulence velocity was
adjusted until both the strong and weak Fe i lines gave the same abundance. And the ionisation equilibrium was used to
derive the surface gravity, and was defined by the equality of the abundances of Fe I and Fe II.

Uncertainties in the effective temperature and micro-turbulence velocity were inferred from the uncertainties in the
slopes of the FeI abundances versus potential excitation and abundance of Fe I versus reduced EW,
respectively. On the other hand, the uncertainty in the surface gravity was obtained  by varying this parameter iteratively around the first guess value
until surface gravity was obtained, that changes the abundance of Fe II by exactly one standard deviation of the
[FeI/H] mean value. In Table~\ref{t_atmparamB} we show the atmospheric parameters and their respective uncertainties for the red clump stars.

The atmospheric parameters for the stars $\#$15490, $\#$29403 and $\#$35658 were not determined because the spectra of these stars
have low S/N. Despite the low S/N also shown by the spectrum of  star $\#$13964 and $\#$32782, we could derive the atmospheric parameters for
but with large  uncertainty (see Table~\ref{t_atmparamB}). In the analysis of the star
$\#$26606, we faced a problem in obtaining the  micro-turbulence velocity due to the low number of absorption lines
with small EW, causing again considerable uncertainty in the atmospheric parameters.

In Table~\ref{abunda-NaB} we show the abundance ratios ([X/Fe]) for  Na, Al, Mg, Ca, Si, Ti, Cr and Ni for the red clump stars. 
Our chemical abundances were normalised to the solar abundances obtained through a high-resolution FEROS solar spectrum
\citep{mon12a}. In Table~\ref{sun} we list  our solar abundances compared to solar abundance of \citet{gre98} and \citet{asp09}.
The total uncertainty of the abundances for the red clump stars are shown in Table~\ref{error-abun}. The uncertainties of the chemical
abundance associated to the errors of the effective temperature, micro-turbulence velocity and surface gravity were calculated
independently, and then quadratically combined to provide the global abundance uncertainty.

\begin{table*}
\tiny
\caption{Atmospheric parameters from spectroscopy of stars of box B.}
\label{t_atmparamB}
\begin{tabular}{cccccccccc}\hline\hline
ID       & $T_{\rm eff}$ & log~$g$     & $\xi$        & [FeI/H]$\pm\sigma$ (\#) & [FeII/H]$\pm\sigma$ (\#) & $\langle$RV$\rangle$ & $E(V-I)$      & $(V-Mv)_{0}$   &       d         \\
         &    (K)        & (dex)       & km\,s$^{-1}$ &                         &                          & (km s$^{-1}$)        &               &                &       (pc)      \\\hline 
11029    & 5250$\pm$200  & 3.3$\pm$0.2 &  2.2$\pm$0.4 &   $-$0.03$\pm$0.13(43)  &  $-$0.03$\pm$0.12(8)     &  76.2$\pm$1.4        & 0.53$\pm$0.19 & 12.41$\pm$0.61 &  3000$\pm$900   \\
13540    & 5300$\pm$200  & 3.1$\pm$0.2 &  1.8$\pm$0.3 &      0.03$\pm$0.14(27)  &     0.05$\pm$0.07(3)     &  74.6$\pm$1.8        & 0.56$\pm$0.11 & 13.62$\pm$0.80 &  5300$\pm$2000  \\
13964$^a$& 4700$\pm$300  & 2.0$\pm$0.4 &  2.5$\pm$0.6 &   $-$0.68$\pm$0.18(23)  &  $-$0.70$\pm$0.18(3)     &  73.9$\pm$2.1        & 0.39$\pm$0.19 & 16.55$\pm$1.17 & 20500$\pm$11600 \\
26606$^b$& 4600$\pm$450  & 2.7$\pm$0.5 &  2.5$\pm$1.0 &   $-$0.46$\pm$0.34(19)  &  $-$0.45$\pm$0.26(3)     &  28.1$\pm$3.6        & 0.30$\pm$0.25 & 13.04$\pm$1.09 &  4000$\pm$2100  \\
27955    & 5250$\pm$250  & 3.8$\pm$0.2 &  2.2$\pm$0.4 &   $-$0.24$\pm$0.15(35)  &  $-$0.24$\pm$0.08(6)     & 116.5$\pm$1.3        & 0.53$\pm$0.10 & 10.46$\pm$0.57 &  1200$\pm$300   \\
28064    & 4700$\pm$150  & 2.1$\pm$0.3 &  2.3$\pm$0.4 &   $-$0.58$\pm$0.14(33)  &  $-$0.57$\pm$0.15(5)     &  68.0$\pm$3.2        & 0.37$\pm$0.15 & 15.45$\pm$0.89 & 12300$\pm$5200  \\
31364    & 5050$\pm$150  & 2.0$\pm$0.2 &  1.5$\pm$0.3 &   $-$0.24$\pm$0.13(31)  &  $-$0.23$\pm$0.08(3)     &  65.6$\pm$0.3        & 0.44$\pm$0.09 & 16.81$\pm$0.90 & 23000$\pm$980   \\
32782    & 5000$\pm$250  & 3.1$\pm$0.3 &  3.0$\pm$0.6 &   $-$0.39$\pm$0.13(16)  &  $-$0.41(1)              &   0.9$\pm$3.0        & 0.49$\pm$0.14 & 12.79$\pm$0.48 &  3600$\pm$800   \\\hline
\end{tabular}

\par \textbf{Notes.} For [Fe I/H] and [Fe II/H], we also show the standard deviation and the number of lines ($\#$) employed. a: 
Large uncertainty in the atmospheric parameters and metallicity of this star due to low S/N. b: Problem in obtaining of 
micro-turbulent velocity ($\xi$) due to the low number of absorption lines with small equivalent width causing considerable
uncertainty in the metallicity.
\end{table*}

\begin{table*}
\tiny
\caption{Abundance Ratios ($[X/Fe]$) for the elements from Na to Cr for the stars from box B.}
\tabcolsep 0.13truecm
\label{abunda-NaB}
\begin{tabular}{lcccccccc}
\hline\hline
\multicolumn{9}{c}{Box B}\\\hline
ID             & [Na/Fe]NLTE         & [Mg/Fe]             & [Al/Fe]             & [Si/Fe]            & [Ca/Fe]            & [Ti/Fe]            & [Cr/Fe]            &  [Ni/Fe]           \\\hline
11029          & $+$0.19(1)          & $+$0.18(1)          &     ---             & $+$0.13$\pm$0.05(2)& $-$0.20$\pm$0.12(4)& $+$0.29$\pm$0.10(5)& $-$0.09$\pm$0.02(3)&$-$0.08$\pm$0.09(10)\\
13540          & $+$0.27$\pm$0.01(2) & ---                 & $+$0.14$\pm$0.04(2) & $+$0.12$\pm$0.16(3)& $+$0.18$\pm$0.11(3)& $+$0.29$\pm$0.04(3)& $-$0.21$\pm$0.10(3)&$+$0.08$\pm$0.13(10)\\
13964          & $+$0.40(1)          & $+$0.66(1)          & $+$0.37$\pm$0.01(2) &     ---            & $+$0.47$\pm$0.07(4)& $+$0.21$\pm$0.14(6)& $+$0.35(1)         & $+$0.24$\pm$0.14(4)\\
26606          & ---                 & $+$0.41$\pm$0.06(2) & $+$0.47(1)          & $+$0.54(1)         &      ---           & $+$0.23$\pm$0.03(2)& $+$0.12(1)         & $+$0.01(1)         \\
27955          & ---                 & $+$0.00$\pm$0.13(2) &      ---            & $+$0.11$\pm$0.10(2)& $-$0.17$\pm$0.17(4)& $+$0.37$\pm$0.08(5)& $-$0.11$\pm$0.03(3)&$+$0.17$\pm$0.16(11)\\
28064          & $+$0.38(1)          & $+$0.24$\pm$0.08(2) & $+$0.10$\pm$0.11(4) & $+$0.12(1)         & $-$0.30$\pm$0.03(2)& +0.22$\pm$0.13(4)  & $+$0.10$\pm$0.13(2)& $-$0.22$\pm$0.07(7)\\
31364          & $+$0.37(1)          & $-$0.21$\pm$0.10(2) &     ---             & $+$0.07$\pm$0.17(3)& $+$0.03$\pm$0.11(3)& $+$0.31$\pm$0.15(3)& $+$0.07$\pm$0.02(2)&$-$0.11$\pm$0.16(10)\\
32782          & $+$0.42(1)          & $+$0.67(1)          & $+$0.59(1)          &    ---             &      ---           & $+$0.54$\pm$0.14(3)& $+$0.10(1)         & $-$0.02$\pm$0.15(5)\\\hline
\hline
\end{tabular}
\par \textbf{Notes.} For all abundances ratios, we also show the standard deviation and the number of lines employed. [Na/Fe] accounts for the NLTE effects calculated as in \citet{gra99}, see text. 
\end{table*}

\begin{table*}
\tiny
\caption{Abundance uncertainties for star from box B.} 
\tabcolsep 0.09truecm
\label{error-abun}
\begin{tabular}{lcccc|cccc}\\\hline\hline
& \multicolumn{4}{c}{11029}& \multicolumn{4}{c}{13540}\\\hline\hline
Element & $\Delta T_{eff}$ & $\Delta\log g$ & $\Delta\xi$ & $\left( \sum \sigma^2 \right)^{1/2}$ & $\Delta T_{eff}$ & $\Delta\log g$ & $\Delta\xi$ & $\left( \sum \sigma^2 \right)^{1/2}$ \\
$_{\rule{0pt}{8pt}}$ & $+$200~K & $+$0.2 & $+$0.4 km\,s$^{-1}$ &  & $+$200~K & $+$0.2 & $+$0.3 km\,s$^{-1}$ & \\
\hline     
Fe\,{\sc i}    & $+$0.13  &    0.00 & $-$0.11 & 0.17 & $+$0.15  & $-$0.01 & $-$0.13 & 0.20\\ 
Fe\,{\sc ii}   & $-$0.17  & $+$0.08 & $-$0.11 & 0.22 & $-$0.15  & $+$0.10 & $-$0.12 & 0.22\\
Na\,{\sc i}    & $+$0.13  &    0.00 & $-$0.06 & 0.14 & $+$0.14  & $-$0.01 & $-$0.07 & 0.16\\
Mg\,{\sc i}    & $+$0.13  & $-$0.02 & $-$0.16 & 0.21 & ---      & ---     & ---     &---  \\
Al\,{\sc i}    & ---      & ---     & ---     & ---  & $+$0.08  & $-$0.01 & $-$0.04 & 0.09\\
Si\,{\sc i}    & $-$0.04  & $+$0.02 & $-$0.04 & 0.06 & $+$0.01  & $+$0.02 & $-$0.06 & 0.06\\
Ca\,{\sc i}    & $+$0.18  & $-$0.02 & $-$0.15 & 0.24 & $+$0.18  & $-$0.02 & $-$0.16 & 0.24\\
Ti\,{\sc i}    & $+$0.25  &    0.00 & $-$0.11 & 0.27 & $+$0.23  & $-$0.01 & $-$0.11 & 0.26\\
Cr\,{\sc i}    & $+$0.25  & $-$0.01 & $-$0.20 & 0.32 & $+$0.24  & $-$0.01 & $-$0.12 & 0.27\\
Ni\,{\sc i}    & $+$0.09  & $+$0.02 & $-$0.12 & 0.15 & $+$0.12  & $+$0.01 & $-$0.13 & 0.18\\
\hline\hline
& \multicolumn{4}{c}{13964}& \multicolumn{4}{c}{26606}\\\hline\hline
Element & $\Delta T_{eff}$ & $\Delta\log g$ & $\Delta\xi$ & $\left( \sum \sigma^2 \right)^{1/2}$ & $\Delta T_{eff}$ & $\Delta\log g$ & $\Delta\xi$ & $\left( \sum \sigma^2 \right)^{1/2}$\\
$_{\rule{0pt}{8pt}}$ & $+$300~K & $+$0.4 & $+$0.6 km\,s$^{-1}$ &  & $+$450~K & $+$0.5 & $+$1.0 km\,s$^{-1}$ &  \\
\hline     
Fe\,{\sc i}  & $+$0.18  & $-$0.03 & $-$0.16 & 0.24& $+$0.34  & $-$0.06 & $-$0.35 & 0.49\\  
Fe\,{\sc ii} & $-$0.35  & $+$0.22 & $-$0.10 & 0.43& $-$0.36  & $+$0.23 & $-$0.17 & 0.46\\  
Na\,{\sc i}  & $+$0.26  & $-$0.01 & $-$0.09 & 0.28& ---      & ---     & ---     & --  \\  
Mg\,{\sc i}  & $+$0.07  & $-$0.02 & $-$0.12 & 0.14& $+$0.12  &    0.00 & $-$0.16 & 0.20\\  
Al\,{\sc i}  & $+$0.15  & $-$0.01 & $-$0.04 & 0.16& $+$0.32  & $-$0.01 & $-$0.16 & 0.36\\  
Si\,{\sc i}  &    ---   &  ---    & ---     &---  & $+$0.24  & $+$0.09 & $-$0.19 & 0.32\\  
Ca\,{\sc i}  & $+$0.35  & $-$0.01 & $-$0.25 & 0.43& ---      & ---     & ---     & --- \\  
Ti\,{\sc i}  & $+$0.50  & $-$0.01 & $-$0.13 & 0.52& $+$0.68  &    0.00 & $-$0.59 & 0.90\\  
Cr\,{\sc i}  & $+$0.28  & $-$0.01 & $-$0.08 & 0.29& $+$0.36  &    0.00 & $-$0.10 & 0.37\\  
Ni\,{\sc i}  & $+$0.05  & $+$0.06 & $-$0.09 & 0.12& $+$0.13  & $+$0.10 & $-$0.33 & 0.37\\  
\hline\hline
& \multicolumn{4}{c}{27955}& \multicolumn{4}{c}{28064}\\\hline\hline
Element & $\Delta T_{eff}$ & $\Delta\log g$ & $\Delta\xi$ & $\left( \sum \sigma^2 \right)^{1/2}$ & $\Delta T_{eff}$ & $\Delta\log g$ & $\Delta\xi$ & $\left( \sum \sigma^2 \right)^{1/2}$ \\
$_{\rule{0pt}{8pt}}$ & $+$250~K & $+$0.2 & $+$0.4 km\,s$^{-1}$ &  & $+$150~K & $+$0.3 & $+$0.4 km\,s$^{-1}$ & \\
\hline     
Fe\,{\sc i}    & $+$0.14  &    0.00 & $-$0.11 & 0.18 & $+$0.14  & $+$0.03 & $-$0.13 & 0.19\\ 
Fe\,{\sc ii}   & $-$0.18  & $+$0.10 & $-$0.08 & 0.22 & $-$0.11  & $+$0.16 & $-$0.07 & 0.21\\
Na\,{\sc i}    & ---      & ---     & ---     & ---  & $+$0.12  & $-$0.01 & $-$0.05 & 0.13\\
Mg\,{\sc i}    & $+$0.10  & $-$0.02 & $-$0.07 & 0.12 & $+$0.05  &    0.00 & $-$0.07 & 0.09\\
Al\,{\sc i}    & ---      & ---     & ---     & ---  & $+$0.08  & $-$0.01 & $-$0.04 & 0.09\\
Si\,{\sc i}    & $+$0.06  & $+$0.03 & $-$0.04 & 0.08 & $-$0.05  & $+$0.07 & $-$0.09 & 0.12\\
Ca\,{\sc i}    & $+$0.22  & $-$0.02 & $-$0.12 & 0.25 & $+$0.18  & $-$0.01 & $-$0.21 & 0.28\\
Ti\,{\sc i}    & $+$0.30  & $-$0.01 & $-$0.15 & 0.34 & $+$0.23  & $-$0.01 & $-$0.09 & 0.25\\
Cr\,{\sc i}    & $+$0.24  & $-$0.01 & $-$0.10 & 0.26 & $+$0.15  & $-$0.01 & $-$0.10 & 0.18\\
Ni\,{\sc i}    & $+$0.10  & $+$0.02 & $-$0.12 & 0.16 & $+$0.10  & $+$0.05 & $-$0.18 & 0.21\\
\hline\hline
& \multicolumn{4}{c}{31364}& \multicolumn{4}{c}{32782}\\\hline\hline
Element & $\Delta T_{eff}$ & $\Delta\log g$ & $\Delta\xi$ & $\left( \sum \sigma^2 \right)^{1/2}$ & $\Delta T_{eff}$ & $\Delta\log g$ & $\Delta\xi$ & $\left( \sum \sigma^2 \right)^{1/2}$ \\
$_{\rule{0pt}{8pt}}$ & $+$150~K & $+$0.2 & $+$0.3 km\,s$^{-1}$ &  & $+$250~K & $+$0.3 & $+$0.6 km\,s$^{-1}$ &  \\
\hline     
Fe\,{\sc i}    & $+$0.14  &    0.00 & $-$0.12 & 0.18& $+$0.15  &   0.00 & $-$0.14 & 0.21\\
Fe\,{\sc ii}   & $-$0.12  & $+$0.13 & $-$0.13 & 0.22& $-$0.21  &$+$0.13 & $-$0.09 & 0.26\\
Na\,{\sc i}    & $+$0.11  & $-$0.01 & $-$0.03 & 0.11& $+$0.19  &$-$0.01 & $-$0.08 & 0.21\\
Mg\,{\sc i}    & $+$0.05  &    0.00 & $-$0.05 & 0.07& $+$0.04  &$-$0.06 & $-$0.10 & 0.12\\
Al\,{\sc i}    & ---      & ---     & ---     & --- & $+$0.11  &$-$0.01 & $-$0.06 & 0.13\\
Si\,{\sc i}    & $+$0.01  & $+$0.03 & $-$0.04 & 0.05&    ---   &   ---  & ---     & --- \\
Ca\,{\sc i}    & $+$0.16  & $-$0.01 & $-$0.17 & 0.23& ---      &   ---  & ---     & --- \\
Ti\,{\sc i}    & $+$0.22  & $-$0.01 & $-$0.09 & 0.24& $+$0.34  &   0.00 & $-$0.11 & 0.36\\
Cr\,{\sc i}    & $+$0.25  & $-$0.02 & $-$0.17 & 0.30& $+$0.36  &   0.00 & $-$0.10 & 0.37\\
Ni\,{\sc i}    & $+$0.14  & $+$0.02 & $-$0.12 & 0.19& $+$0.11  &$+$0.04 & $-$0.15 & 0.19\\
\hline\hline\\
\end{tabular}

\par \textbf{Notes.} Each column gives the variation of the abundance caused by the variation in $T_{\rm eff}$, $\log g$ and $\xi$. The last column for each star gives the compounded rms uncertainty of the second to fourth columns.

\end{table*}

\begin{table}
\caption{Observed Fe\,{\sc i} and Fe\,{\sc ii} lines.}
\scriptsize
\begin{tabular}{cccccccccccc}\\\hline\hline
\label{tabelFea}
        &            &         &          & \multicolumn{8}{c}{Equivalent Widths (m\AA)}
\\\hline
        &            &         &          & \multicolumn{8}{c}{Star} \\
\cline{5 - 12}
Element &  $\lambda$\,(\AA) & $\chi$(eV) & log $gf$  & 11029 & 13540 & 13964 & 26606 & 27955 & 28064 & 31364 & 32782  \\
\hline
Fe\,{\sc i} &   5162.27  &  4.18  &   0.079 &  --- & 153 & --- & --- &  --- & --- & 143& ---\\
 &   5198.71  &  2.22  &  -2.140 &  --- & --- & ---& --- & --- & --- & 126& ---\\
 &   5242.49  &  3.63  &  -0.970 &  --- & --- & ---& --- &  ---& 126 &  98& ---\\
 &   5288.52  &  3.69  &  -1.510 &  --- & --- & ---& --- &  ---& --- &  82& ---\\
 &   5307.36  &  1.61  &  -2.970 &  160 & 140 & ---& ---&  149 & --- & ---& ---\\
 &   5315.05  &  4.37  &  -1.400 &   64 &  63 & ---& ---&  --- & --- & ---& ---\\
 &   5321.11  &  4.43  &  -1.190 &  --- & --- & ---& ---&  --- &  69 & ---& ---\\
 &   5322.04  &  2.28  &  -2.840 &  --- & --- & 128& 127&  108 & --- & ---& ---\\
 &   5364.87  &  4.45  &   0.230 &  --- & 153 & ---& ---&  --- & --- & ---& ---\\
 &   5373.71  &  4.47  &  -0.710 &   95 &  97 &  84& ---&  --- & --- &  84& ---\\
 &   5389.48  &  4.42  &  -0.250 &  159 & --- & ---& ---&  131 & --- & ---& ---\\
 &   5393.17  &  3.24  &  -0.720 &  --- & 156 & ---& ---&  --- & --- & 149& ---\\
 &   5417.03  &  4.42  &  -1.530 &  --- & --- & ---& ---&  --- & --- & ---&  53\\
 &   5441.34  &  4.31  &  -1.580 &   45 & --- & ---& ---&  --- &  42 & ---& ---\\
 &   5445.04  &  4.39  &  -0.041 &  --- & 133 & ---& ---&  --- & 153 & ---& ---\\
 &   5522.45  &  4.21  &  -1.400 &  70  & --- & ---&  39&  --- & --- & ---& ---\\
 &   5531.98  &  4.91  &  -1.460 &  --- & --- & ---&  59&   20 &  30 & ---& ---\\
 &   5532.75  &  3.57  &  -2.000 &  --- & --- &  40& ---&  --- & --- & ---& ---\\
 &   5554.90  &  4.55  &  -0.380 &  --- & --- & ---& ---&  125 & --- & ---& ---\\
 &   5560.21  &  4.43  &  -1.040 &  85  & --- & ---&  67&  --- & --- &  70& ---\\
 &   5567.39  &  2.61  &  -2.560 &  102 & --- & ---& ---&  --- & 124 &  83& ---\\
 &   5584.77  &  3.57  &  -2.170 &  --- & --- & ---& ---&  --- & --- &  68& ---\\
 &   5624.02  &  4.39  &  -1.330 &   78 & --- & ---& ---&  --- & --- & ---& ---\\ 
 &   5633.95  &  4.99  &  -0.120 &  --- & --- & 105& 107&   76 & --- & ---& ---\\
 &   5635.82  &  4.26  &  -1.740 &  56  & --- & ---& ---&   25 & --- & ---&  40\\
 &   5638.26  &  4.22  &  -0.720 &  116 & --- & ---& ---&  --- & --- & ---& 131\\
 &   5686.53  &  4.55  &  -0.450 &  122 & --- & ---& ---&  --- & --- & ---& ---\\ 
 &   5691.50  &  4.30  &  -1.370 &  --- & --- & ---& ---&  64  &  83 & ---& ---\\
 &   5705.47  &  4.30  &  -1.360 &  --- & --- &  82& ---&   49 &  67 &  51&  76\\
 &   5717.83  &  4.28  &  -0.979 &  --- & --- & ---&  88&   80 & --- & ---& ---\\
 &   5731.76  &  4.26  &  -1.150 &  101 & --- & ---& ---&  --- &  91 & ---& ---\\
 &   5806.73  &  4.61  &  -0.900 &  75  & --- & ---& ---&  --- & --- & ---&  80\\
 &   5814.81  &  4.28  &  -1.820 &  --- & --- & ---& ---&  --- &  37 &  29& ---\\ 
 &   5852.22  &  4.55  &  -1.180 &  78  &  76 & ---& ---&  --- & --- &  51& ---\\
 &   5883.82  &  3.96  &  -1.210 &  --- & --- &  95& ---&  79  & 101 & ---& ---\\
 &   5916.25  &  2.45  &  -2.990 &  --- &  99 &  97& ---& ---  & 108 &  86&  99\\
 &   5934.65  &  3.93  &  -1.020 &  111 & 103 & 105& ---& 113  & --- & ---& ---\\
 &   6020.17  &  4.61  &  -0.210 &  --- & --- & ---& ---&  128 & --- & ---& ---\\ 
 &   6024.06  &  4.55  &  -0.060 &  141 & --- & 124& ---&  128 & 123 & ---& ---\\
 &   6027.05  &  4.08  &  -1.090 &  --- & 107 & 112&  92&   87 & --- & ---& ---\\
 &   6056.01  &  4.73  &  -0.400 &  --- & --- &  92& ---&   99 & --- &  80& 127\\
 &   6079.01  &  4.65  &  -0.970 &  76  & --- &  60& ---&  --- & --- & ---& ---\\
 &   6082.71  &  2.22  &  -3.580 &  --- & --- & ---& ---&  --- &  98 & ---& ---\\ 
 &   6093.64  &  4.61  &  -1.350 &  46  & --- & ---& ---&  --- & --- & ---& ---\\
 &   6096.66  &  3.98  &  -1.780 &  70  &  65 & ---& ---&  --- & --- &  65& ---\\
 &   6120.25  &  0.91  &  -5.950 &  --- & --- & ---& ---&  --- &  42 & ---& ---\\ 
 &   6151.62  &  2.18  &  -3.290 &  98  &  92 & 112& ---&   79 & 103 & ---& ---\\
 &   6157.73  &  4.08  &  -1.110 & ---  & --- & ---&  85&  --- & --- &  85& ---\\
 &   6165.36  &  4.14  &  -1.470 &  64  &  66 &  75&  93&   78 & --- &  71& ---\\
 &   6170.51  &  4.79  &  -0.380 & ---  & 102 & ---& ---&  --- & --- & ---& ---\\
 &   6173.34  &  2.22  &  -2.880 & 126  & --- & ---& ---& ---  & 146 & ---& 155\\
 &   6187.99  &  3.94  &  -1.570 &  82  &  77 &  89& ---&  57  &  88 &  72& ---\\
 &   6200.31  &  2.60  &  -2.440 &  130 & --- & 133& ---&  118 & 144 & ---& ---\\
\hline
\end{tabular}
\end{table}
 
\begin{table}
\noindent
\scriptsize
\begin{tabular}{cccccccccccc}
\multicolumn{12}{c}{Table 6, continued}\\
\hline\hline
 &            &         &          & \multicolumn{8}{c}{Equivalent Widths (m\AA)}
\\\hline
        &            &         &          & \multicolumn{8}{c}{Star} \\
\cline{5 - 12}
Element &  $\lambda$\,(\AA) & $\chi$(eV) & log $gf$  & 11029 & 13540 & 13964 & 26606 & 27955 & 28064 & 31364 & 32782  \\
\hline
Fe\,{\sc i} &   6213.43  &  2.22  &  -2.480 &  149 & --- & ---& ---&  --- & --- & ---& ---\\
 &   6254.26  &  2.28  &  -2.440 &  157 & --- & ---& ---&  --- & --- & ---& ---\\
 &   6265.13  &  2.18  &  -2.550 &  147 & --- & ---& ---&  143 & --- & 124& ---\\
 &   6322.69  &  2.59  &  -2.430 & 137  & --- & ---& ---&  124 & --- & ---& 142\\
 &   6380.74  &  4.19  &  -1.320 &  86  &  84 & ---& ---&   83 &  81 & ---& ---\\
 &   6392.54  &  2.28  &  -4.030 &  62  & --- & ---& ---&  35  &  51 & ---& ---\\
 &   6411.65  &  3.65  &  -0.660 &  --- & 139 & ---& ---&  --- & --- & ---& ---\\
 &   6421.35  &  2.28  &  -2.010 &  --- & --- & ---& ---&  149 & --- & 151& ---\\
 &   6436.41  &  4.19  &  -2.460 &  --- & --- & ---& ---&   20 &  10 & ---& ---\\
 &   6469.19  &  4.83  &  -0.620 &  98  & 79  & ---& 116&  75  &  72 & ---& ---\\
 &   6574.23  &  0.99  &  -5.020 &  --- &  86 & ---& 101&  61  & 102 &  78& ---\\ 
 &   6591.31  &  4.59  &  -2.070 &  20  & --- & ---& ---&  --- & --- & ---& ---\\
 &   6593.87  &  2.44  &  -2.420 &  --- & --- & ---& ---& 129  & --- & ---& ---\\ 
 &   6597.56  &  4.79  &  -0.920 &  60  & --- & ---&  80& ---  & --- &  46& ---\\
 &   6608.03  &  2.28  &  -4.030 &  61  &  37 &  58&  90& ---  &  63 & ---& ---\\
 &   6609.11  &  2.56  &  -2.690 & 122  & --- & ---& ---& ---  & 110 & ---& 133\\
 &   6646.93  &  2.61  &  -3.990 & ---  &  23 & ---& ---&  18  &  25 & ---& ---\\
 &   6653.85  &  4.14  &  -2.520 &  --- & --- & ---& ---& ---  & --- &  22& ---\\ 
 &   6703.57  &  2.76  &  -3.160 &  68  &  79 &  78&  53&  41  & --- & ---&  55\\
 &   6710.32  &  1.80  &  -4.880 & ---  & --- & ---&  68&  --- & --- & ---& ---\\
 &   6739.52  &  1.56  &  -4.950 &   39 & --- &  57&  86&  --- &  58 &  34&  59\\
 &   6750.15  &  2.42  &  -2.620 &  118 & --- & ---& 147&  --- & --- & 118& 148\\
 &   6752.71  &  4.64  &  -1.200 &  59  &  47 & ---& ---&   45 & --- &  60& ---\\
 &   6806.85  &  2.73  &  -3.210 &  77  & --- &  97& 113&  --- &  83 & ---& ---\\
 &   6820.37  &  4.64  &  -1.170 & ---  &  74 & ---& ---&  --- &  43 &  50& ---\\
 &   6841.34  &  4.61  &  -0.600 & 124  & --- & ---& ---&  --- & --- & ---& ---\\
 &   6851.64  &  1.61  &  -5.320 & ---  & --- & ---&  52&  --- & --- &  24& ---\\
 &   6858.15  &  4.61  &  -0.930 & ---  &  91 &  75& ---&   88 &  81 & ---&  64\\
 &   7130.92  &  4.22  &  -0.700 &  129 & 126 & 126& ---&  111 & 115 & 113& ---\\
 &   7132.99  &  4.08  &  -1.610 &  --- & --- &  76& ---&  --- & --- &  66& ---\\
Fe\,{\sc ii} &  5132.66  &  2.81  &  -4.000 & --- & --- & ---& ---&  26 & --- & ---& ---\\
  &  5234.62  &  3.22  &  -2.240 &  --- & --- & ---& ---& 108 & --- & ---& ---\\  
  &  5425.25  &  3.20  &  -3.210 &  60 &  70 &  74&  39&  44 & --- &  75&  59 \\
  &  5534.83  &  3.25  &  -2.770 &  95 & --- & ---& ---&  76 & --- & ---& --- \\  
  &  5991.37  &  3.15  &  -3.560 &  66 & --- & ---& ---& --- & --- & ---& --- \\
  &  6084.10  &  3.20  &  -3.800 &  39 & --- & ---& ---&  17 &  26 & ---& --- \\
  &  6149.25  &  3.89  &  -2.720 &  53 & --- &  49&  45& --- &  42 &  70& --- \\
  &  6247.55  &  3.89  &  -2.340 &  82 &  86 &  56& ---& --- &  78 &  81& --- \\
  &  6369.46  &  2.89  &  -4.110 &  ---& --- & ---& ---& --- &  43 & ---& --- \\ 
  &  6416.92  &  3.89  &  -2.680 &  63 &  67 & ---& ---&  43 &  41 & ---& --- \\
  &  6432.68  &  2.89  &  -3.580 &  61 & --- & ---&  36&  ---& --- & ---& --- \\
\hline
\end{tabular}
\end{table}

\begin{table*}
\caption{Other lines studied.}
\scriptsize
\begin{tabular}{ccccccccccccc}
\label{tabellinesa}
\\\hline\hline
    & & & & &\multicolumn{8}{c}{Equivalent Widths (m\AA)} \\\hline
\multicolumn{5}{c}{} & \multicolumn{8}{c}{Star}\\
\cline{6-13}
Element & $\lambda$ & $\chi$(eV) & $\log gf$ & Ref & 11029 & 13540 & 13964 & 26606 & 27955 & 28064 & 31364 & 32782\\
\hline
Na\,{\sc i} & 6154.22 & 2.10 & $-$1.51 & PS   &  73 &  79 & --- & --- & --- &  66 &  59 &  88 \\
Na\,{\sc i} & 6160.75 & 2.10 & $-$1.21 & R03  & --- &  99 &  89 & --- & --- & --- & --- & --- \\
Mg\,{\sc i} & 4730.04 & 4.34 & $-$2.39 & R03  &  81 & --- & --- & --- & --- & --- & --- & --- \\ 
Mg\,{\sc i} & 5711.10 & 4.34 & $-$1.75 & R99  & 146 & --- & --- & --- & 127 & --- & --- & --- \\
Mg\,{\sc i} & 7387.70 & 5.75 & $-$0.87 & MR94 & --- & --- & 115 & 120 &  77 & --- &  54 & --- \\
Mg\,{\sc i} & 8717.83 & 5.91 & $-$0.97 & WSM  & --- & --- & --- &  88 & --- &  70 & --- & 129 \\
Mg\,{\sc i} & 8736.04 & 5.94 & $-$0.34 & WSM  & --- & --- & --- & --- & --- & 112 &  91 & --- \\
Al\,{\sc i} & 6698.67 & 3.14 & $-$1.63 & R03  & --- & --- & --- &  74 & --- & --- & --- & --- \\
Al\,{\sc i} & 7835.32 & 4.04 & $-$0.58 & R03  & --- &  62 &  45 & --- & --- &  32 & --- &  82 \\
Al\,{\sc i} & 7836.13 & 4.02 & $-$0.40 & R03  & --- &  72 &  59 & --- & --- &  54 & --- & --- \\
Al\,{\sc i} & 8772.88 & 4.02 & $-$0.25 & R03  & --- & --- & --- & --- & --- &  66 & --- & --- \\
Al\,{\sc i} & 8773.91 & 4.02 & $-$0.07 & R03  & --- & --- & --- & --- & --- &  65 & --- & --- \\
Si\,{\sc i} & 5793.08 & 4.93 & $-$2.06 &  R03 & --- &  76 & --- & --- &  51 & --- & --- & --- \\
Si\,{\sc i} & 6125.03 & 5.61 & $-$1.54 &  E93 &  55 &  56 & --- & --- & --- & --- &  40 & --- \\
Si\,{\sc i} & 6131.58 & 5.62 & $-$1.69 &  E93 & --- & --- & --- & --- & --- & --- &  45 & --- \\
Si\,{\sc i} & 6145.02 & 5.61 & $-$1.43 &  E93 &  58 & --- & --- & --- & --- & --- & --- & --- \\
Si\,{\sc i} & 6155.14 & 5.62 & $-$0.77 &  E93 & --- &  95 & --- & --- &  86 & --- &  79 & --- \\
Si\,{\sc i} & 8728.01 & 6.18 & $-$0.36 &  E93 & --- & --- & --- &  97 & --- & --- & --- & --- \\
Si\,{\sc i} & 8742.45 & 5.87 & $-$0.51 &  E93 & --- & --- & --- & --- & --- &  86 & --- & --- \\
Ca\,{\sc i} & 6161.30 & 2.52 & $-$1.27 &  E93 & --- & 107 & 125 & --- & --- & --- & --- & --- \\
Ca\,{\sc i} & 6166.44 & 2.52 & $-$1.14 &  R03 &  87 & 112 & 140 & --- & 101 & --- &  89 & --- \\
Ca\,{\sc i} & 6169.04 & 2.52 & $-$0.80 &  R03 & 129 & --- & 154 & --- & --- & --- & --- & --- \\
Ca\,{\sc i} & 6169.56 & 2.53 & $-$0.48 & DS91 & 142 & --- & --- & --- & --- & --- & --- & --- \\
Ca\,{\sc i} & 6455.60 & 2.51 & $-$1.29 &  R03 &  78 & --- & 121 & --- &  64 & --- & --- & --- \\
Ca\,{\sc i} & 6471.66 & 2.51 & $-$0.69 &  S86 & --- & 132 & --- & --- & 115 & 117 & 106 & --- \\
Ti\,{\sc i} & 4758.12 & 2.25 &    0.420 & MFK & --- &  92 & --- & 125 & --- & --- & --- & --- \\

Ti\,{\sc i} & 5039.96 & 0.02 & $-$1.130 & MFK & --- & --- & --- & --- & 148 & --- & --- & --- \\
Ti\,{\sc i} & 5043.59 & 0.84 & $-$1.733 & MFK &  63 & --- & --- & --- & --- & --- & --- & --- \\
Ti\,{\sc i} & 5062.10 & 2.16 & $-$0.464 & MFK & --- & --- & --- & --- &  50 & --- & --- & --- \\
Ti\,{\sc i} & 5113.45 & 1.44 & $-$0.880 & E93 & --- & --- & --- & --- & --- &  93 & --- & --- \\
Ti\,{\sc i} & 5223.63 & 2.09 & $-$0.559 & MFK & --- & --- &  50 & --- & --- & --- & --- & --- \\
Ti\,{\sc i} & 5295.78 & 1.05 & $-$1.633 & MFK & --- & --- & --- & --- & --- &  72 & --- & --- \\
Ti\,{\sc i} & 5490.16 & 1.46 & $-$0.937 & MFK & --- &  69 & --- & --- & --- & --- & --- & --- \\
Ti\,{\sc i} & 5689.48 & 2.30 & $-$0.469 & MFK & --- & --- & --- & --- & --- &  46 & --- &  64 \\
Ti\,{\sc i} & 5866.46 & 1.07 & $-$0.871 & E93 & 106 & --- & --- & --- & 109 & --- &  95 & --- \\
Ti\,{\sc i} & 5922.12 & 1.05 & $-$1.465 & MFK & --- & --- &  83 & --- & --- & --- &  76 & --- \\
Ti\,{\sc i} & 5978.55 & 1.87 & $-$0.496 & MFK &  81 &  75 & --- & --- &  75 & --- & --- & --- \\
Ti\,{\sc i} & 6091.18 & 2.27 & $-$0.370 & R03 &  67 & --- & --- & --- & --- & --- &  42 & --- \\
Ti\,{\sc i} & 6126.22 & 1.07 & $-$1.370 & R03 &  84 & --- & 103 & 128 & --- &  81 & --- & 101 \\
Ti\,{\sc i} & 6258.11 & 1.44 & $-$0.355 & MFK & --- & --- & 118 & --- & --- & --- & --- & --- \\
Ti\,{\sc i} & 6261.11 & 1.43 & $-$0.480 & B86 & --- & --- & 127 & --- & 105 & --- & --- & --- \\
Ti\,{\sc i} & 6554.24 & 1.44 & $-$1.219 & MFK & --- & --- &  49 & --- & --- & --- & --- & 100 \\
Cr\,{\sc i} & 5193.50 & 3.42 & $-$0.720 & MFK &  27 & --- & --- & --- & --- & --- & --- & --- \\
Cr\,{\sc i} & 5214.13 & 3.37 & $-$0.740 & MFK & --- & --- & --- & --- &  20 & --- & --- & --- \\
Cr\,{\sc i} & 5296.70 & 0.98 & $-$1.390 & GS  & 154 & --- & --- & --- & --- & --- & --- & --- \\
Cr\,{\sc i} & 5304.18 & 3.46 & $-$0.692 & MFK & --- & --- & --- &  42 & --- & --- & --- & --- \\
Cr\,{\sc i} & 5345.81 & 1.00 & $-$0.980 & MFK & --- & --- & --- & --- & --- & --- & 151 & --- \\
Cr\,{\sc i} & 5348.32 & 1.00 & $-$1.290 & GS  & 158 & 132 & --- & --- & 144 & --- & --- & --- \\
Cr\,{\sc i} & 5783.07 & 3.32 & $-$0.500 & MFK & --- & --- &  64 & --- &  47 &  73 & --- & --- \\
Cr\,{\sc i} & 5787.93 & 3.32 & $-$0.080 & GS  & --- &  58 & --- & --- & --- &  75 & --- & --- \\
Cr\,{\sc i} & 6330.09 & 0.94 & $-$2.920 & R03 & --- &  57 & --- & --- & --- & --- &  65 &  85 \\
Ni\,{\sc i} & 4904.42 & 3.54 & $-$0.170 & MFK & --- & 125 & --- & --- & 143 & --- & --- & 136 \\
Ni\,{\sc i} & 4935.83 & 3.94 & $-$0.360 & MFK & --- &  83 & --- & --- &  88 &  70 & --- & --- \\
\hline
\end{tabular}
\end{table*}
 
\begin{table*}
\noindent
\scriptsize
\begin{tabular}{ccccccccccccc}
\multicolumn{13}{c}{Table 7, continued.}\\
\\\hline\hline
    & & & & &\multicolumn{8}{c}{Equivalent Widths (m\AA)} \\\hline
\multicolumn{5}{c}{} & \multicolumn{8}{c}{Star}\\
\cline{6-13}
Element & $\lambda$ & $\chi$(eV) & $\log gf$ & Ref & 11029 & 13540 & 13964 & 26606 & 27955 & 28064 & 31364 & 32782\\
\hline
Ni\,{\sc i} & 4953.21 & 3.74 & $-$0.660 & MFK & --- & --- & --- & --- &  88 & --- & --- & --- \\
Ni\,{\sc i} & 5010.94 & 3.63 & $-$0.870 & MFK & --- & --- &  87 & --- & --- & --- & --- & --- \\
Ni\,{\sc i} & 5578.73 & 1.68 & $-$2.640 & MFK & 108 & --- & --- & --- & --- & 110 &  90 & --- \\
Ni\,{\sc i} & 5593.75 & 3.90 & $-$0.840 & MFK &  60 & --- & --- & --- &  73 &  45 & --- &  85 \\
Ni\,{\sc i} & 5643.09 & 4.17 & $-$1.250 & MFK &  24 &  40 & --- & --- & --- & --- & --- & --- \\
Ni\,{\sc i} & 5748.36 & 1.68 & $-$3.260 & MFK & --- & --- & --- & --- & --- & --- & --- &  73 \\
Ni\,{\sc i} & 5805.23 & 4.17 & $-$0.640 & MFK & --- & --- & --- & --- &  72 &  43 & --- & --- \\
Ni\,{\sc i} & 6053.69 & 4.24 & $-$1.070 & MFK & --- &  49 & --- & --- & --- & --- & --- & --- \\
Ni\,{\sc i} & 6086.29 & 4.27 & $-$0.510 & MFK &  62 &  81 & --- & --- & --- &  50 & --- & --- \\
Ni\,{\sc i} & 6128.98 & 1.68 & $-$3.320 & MFK & --- &  78 & --- & --- & 64  &  67 &  57 & --- \\
Ni\,{\sc i} & 6176.82 & 4.09 & $-$0.264 & R03 & --- & --- & --- & --- &  93 & --- & --- & --- \\
Ni\,{\sc i} & 6186.72 & 4.11 & $-$0.960 & MFK &  56 & --- &  47 & --- & --- & --- &  50 & --- \\
Ni\,{\sc i} & 6204.61 & 4.09 & $-$1.150 & MFK &  42 & --- & --- & --- & --- & --- & --- & --- \\
Ni\,{\sc i} & 6223.99 & 4.11 & $-$0.980 & MFK & --- & --- &  57 & --- & --- & --- & --- & --- \\
Ni\,{\sc i} & 6230.10 & 4.11 & $-$1.260 & MFK & --- & --- &  44 & --- & --- & --- &  33 & --- \\
Ni\,{\sc i} & 6327.60 & 1.68 & $-$3.150 & MFK & --- & --- & --- & --- &  85 & --- &  63 & --- \\
Ni\,{\sc i} & 6482.81 & 1.94 & $-$2.630 & MFK & --- & --- & --- & --- &  83 & --- & --- & 100 \\
Ni\,{\sc i} & 6532.88 & 1.94 & $-$3.390 & MFK & --- & --- & --- & --- &  57 &  44 & --- & --- \\
Ni\,{\sc i} & 6586.32 & 1.95 & $-$2.810 & MFK & --- &  84 & --- & --- & --- & --- & --- &  95 \\
Ni\,{\sc i} & 6635.14 & 4.42 & $-$0.830 & MFK & --- &  41 & --- & --- & --- & --- &  40 & --- \\
Ni\,{\sc i} & 6643.64 & 1.68 & $-$2.030 & MFK & 153 & 144 & --- & --- & --- & --- & 129 & --- \\
Ni\,{\sc i} & 6767.78 & 1.83 & $-$2.170 & MFK & 141 & --- & --- & --- & --- & --- & 105 & --- \\
Ni\,{\sc i} & 6772.32 & 3.66 & $-$0.970 & R03 &  73 & --- & --- &  83 & --- &  84 & --- & --- \\
Ni\,{\sc i} & 7788.93 & 1.95 & $-$1.990 & E93 & 151 & 140 & --- & --- & --- & --- & 130 & --- \\
\hline
\\
\end{tabular}
\par References: B86: \citet{bla86}; Ca07: \citet{car07}; D2002: \citet{dep02};
\par DS91: \citet{dra91}; E93: \citet{edv93}; GS: \citet{gra88};
\par MFK: \citet{mar02}; MR94: \citet{mcw94};
\par PS: \citet{pre01}; R03: \citet{red03};
\par R99: \citet{red99}; WSM: \citet{wie69}.

\end{table*}

\begin{table}
\caption{Adopted solar abundances.}
\tabcolsep 0.33truecm
\label{sun}
\begin{tabular}{lccc}\\\hline\hline
Element              & This  & Grevesse \&  & Asplund                \\
$_{\rule{0pt}{8pt}}$ & work  & Sauval (1998)& \textit{et al.} (2009) \\
\hline     
Fe                   & 7.50  & 7.50         & 7.50                   \\ 
Na                   & 6.26  & 6.33         & 6.24                   \\
Mg                   & 7.55  & 7.58         & 7.60                   \\
Al                   & 6.31  & 6.47         & 6.45                   \\
Si                   & 7.61  & 7.55         & 7.51                   \\
Ca                   & 6.37  & 6.36         & 6.34                   \\
Ti                   & 4.93  & 5.02         & 4.95                   \\
Cr                   & 5.65  & 5.67         & 5.64                   \\
Ni                   & 6.29  & 6.25         & 6.22                   \\
\hline\\
\end{tabular}
\end{table}

\subsection{Box~C and D: stellar parameters}
\label{s_CDparams}

The temperature, gravity, and rotational velocity of Box~C stars were measured by fitting the H$_\alpha$ and H$_\beta$ Balmer
lines and the 4922~\AA\ He~I feature with synthetic spectra, as done in \citet{Majaess13} for similar MS stars. To this aim,
we employed the {\it fitprof21} code, developed by \citet{Bergeron92} and \citet{Saffer94}, and subsequently modified by
\citet{Napiwotzki99}. The routine was fed with a grid of solar-metallicity LTE model spectra
(T$_\mathrm{eff}$=8000--30000~K, $\log{g}$=3.5--5.0~dex) generated from ATLAS9 \citep{kur93} model atmospheres through the
Lemke’s version\footnote{http://a400.sternwarte.uni-erlangen.de/\~ai26/linfit/linfor.html} of the LINFOR program (developed
originally by Holweger, Steffen, and Steenbock at Kiel University). In fact, deviations from LTE have negligible effects
on the Balmer and He lines at the temperatures of program stars \citep{Nieva07}. The routine determines the best-fit
parameters through a $\chi^2$ minimisation statistics. Extensive details about the synthetic spectra fitting procedure can be 
found in \citet{moni12}.

While the method ideally works when the whole Balmer series can be simultaneously fit, a minimum of two features is required
to avoid the degeneracy between temperature and gravity. In our case, the profile of the only He line could constrain
the rotational velocity and, to a limited extent, the temperature, but the two Balmer lines were needed for a reliable
determination of T$_\mathrm{eff}$ and $\log{g}$. This was not possible  for three targets, where H$_\beta$ line did not
fall in the spectral range. We could still obtain a good fit of the single H$_\alpha$ line for the stars \#6507 and \#8542,
although the larger errors reflect the high uncertainty of the measurements. For  star \#12018, on the contrary, we had
to assume $\log{g}$=4.2, as typical of a MS star, and adopt only the temperature as fit parameter. 

The strength of Balmer lines has a maximum at $\approx$10000~K, and it declines both for hotter and cooler stars. The temperatures obtained
by one Balmer line only could thus present two acceptable solutions, symmetric with respect to A0 spectral type. We therefore analysed the
results of the three aforementioned stars in more detail. We indeed found a secondary solution for the star \#8542, with a local minimum of
the $\chi^2$ statistics, on the other side of the Balmer minimum at 10100~K. However, this minimum of $\chi^2$ is shallower than the main
solution at 8100~K, which is therefore more likely and should be preferred. We did not find such a secondary minimum for star \#12018, and
the fitting routine converged to the same solution at 8700~K even if it was forced to start from a hot first guess at
$T_\mathrm{Jeff}>12000$~K. On the other hand, a cooler solution for the star \#6507 is not acceptable, because $T_\mathrm{eff}<9000$~K would
return $E(V-I)<0.37$ and a distance $d\approx 10$~kpc, clearly offset from the reddening-distance relation depicted by the other stars in
box~C and D (see below in Fig.~10). In addition, this object is of little interest because most likely not a genuine MS star (see later).
In conclusion, our tests indicate that, despite fitting one Balmer line only, the solutions we find for these three stars are either the only
one acceptable, or the most likely.
To check the effects of
the use of solar metallicity models, we repeated the measurements with models with [Fe/H]=$-0.5$, but we found that the
results changed by less than 0.5$\sigma$ in all the cases. The results are given in Table~\ref{t_paramD}. The resulting
surface gravities indicate that all the targets are MS stars, with the exception of the star \#6507. The measurements for
this star are affected by large errors, so that it could still be considered a MS object within the error bars, but the
high $\log{g}$ value suggests that this could be a foreground sub-dwarf B-type (sdB) star.

The temperature of Box~D targets was determined by fitting the profile of temperature-sensitive lines with synthetic spectra
drawn from the library of \citet{Coelho05}. We adopted the same routines of \citet{moni10}, where detailed information about
the measurement procedure can be found. The main feature for our measurements was the H$_\alpha$ Balmer line, which is
a good indicator of temperature in the range $T_\mathrm{eff}$=5000--6500~K, insensitive to metallicity and surface gravity
\citep{Fuhrmann94}. Its wings were fitted with solar-metallicity templates in steps of 250~K, and the $\chi^2$ 
 was minimised to find the best estimate of the target temperature. We verified that varying the metallicity of the
synthetic templates had only negligible effects on the results.

One gravity-sensitive feature was observable in the spectrum of some targets, either the MgIb triplet \citep{Kuijken89} or
the Na~I doublet at 5890--5893~\AA. However, the low resolution of the data prevented an estimate of $\log{g}$, because
only large mismatches ($>1$~dex) between the template and object line wings could be appreciated. Hence, the targets were
assumed MS stars along the whole process, with a fixed surface gravity of $\log{g}$=4.1~dex, as strongly suggested by their
position in the CMD. We nevertheless confirmed this hypothesis by checking that the profiles of available lines were compatible
with it. On the other hand, this assumption was also confirmed by later distance estimates because, had one of these stars
been either a faint sub-dwarf or a bright giant, its distance would have resulted extremely large or short, respectively,
which is not the case (see Table~\ref{t_paramD}). The MgIb triplet and the Na~I doublet were instead used to derive
independent estimates of $T_\mathrm{eff}$, with a procedure identical to that used for H$_\alpha$. The final results and
their associated errors were obtained from the average and the error-of-the-mean of these measurements.

\subsection{Reddening and distances}

The reddening for the stars of box B were estimate using isochrones of \citet{bert08,bert09}
to obtain the $(V-I)$ intrinsic color of each star. In the Table~\ref{t_atmparamB} we show $E(V-I)$ values obtained for the stars of box~B.
We also calculated the distances for each star of the box~B using the equation:

\begin{eqnarray}
\log d\; ({\rm pc})\; & = & 1/2 [\log \frac{M_{\star}}{M_{\odot}} 
 + 0.4\left(V-A_{ V}+BC\right) \nonumber \\
 & &  {{\,}\atop{\,}} + 4\log T_{\rm eff} - \log g_{\star} - 10.62].
\end{eqnarray}

\noindent
Where $T_{eff}$ and log $g_{\star}$ are the effective temperature and surface gravity, respectively, and $M$ is the mass 
obtained through the evolutionary tracks of \citet{bert08} and \citet{bert09}.
The photometric data of Table~\ref{t_specdata} were combined with bolometric corrections ($BC$) defined by the 
relations of \citet{alo99}. The extinction in $V$ ($A_{V}$) for each star was calculated using the 
reddening $E(V-I)$ shown in the Table~\ref{t_atmparamB}, the non-standard absorption law valid for the third Galactic quadrant, where
$R_V$=2.9 \citep{turner14} and $E(B-V)=0.7955\times E(V-I)$ \citep{turner11}.
For the Sun we adopted $M_{bol \odot} = 4.74$ mag \citep{bes98}, $T_{\rm eff \odot} = 5700$ K and $\log g_\odot = 4.3$ dex.

We also performed a rough estimate of the age of the stars from box B using spectroscopic atmospheric parameters and isochrones of 
\citet{bert08} and \citet{bert09}. We note that such a sample is composed of a great mix of stars with the age ranging
from 1.2~Myr to 10~Gyr. This huge range is not  unexpected when considering a sample containing both  thin and thick disk stars.

The reddening and distance of box~C and D stars were derived similarly. The intrinsic colours and absolute magnitudes of box~C stars
were derived from comparison of their position in the temperature--gravity plane with the same solar-metallicity isochrones used for
box~B. The intrinsic color, compared with the observed one, returned the reddening $E(V-I)$, which was used to derive $A_{V}$ with the
same equations given above. As we had no gravity information for box~D stars, but we argued that they are all MS objects, we adopted for
them the absolute magnitude of solar-metallicity ZAMS objects at the corresponding temperature. From the  derived absolute magnitude,
the observed $V$ magnitude, and $A_{V}$, the distances were straightforwardly computed. We chose to use $(V-I)$ as temperature indicator
for consistency with what done in Box~B. However, a bluer color could be a better choice for Box~C, where the stars are noticeably hotter
than in the other two groups. To test if our choice would alter the results, we repeated the procedure using $(B-V)$ instead of
$(V-I)$ in Box~C. The reddening values thus derived are compatible within errors with those previously obtained, with a mean difference
and standard deviation of $-0.04\pm0.05$~mag, and no clear trend with temperature.

\begin{table*}
\caption{Atmospheric parameters from spectroscopy of box C.}
\label{t_paramC}
\begin{tabular}{cccccccc}\hline\hline
ID     &  $T_{\rm eff}$ & $\log{g}$   & $v sin i$   & $\langle$ RV $\rangle$ & $E(V-I)$      & $(V-Mv)_{0}$   &    d           \\
       &     K          &  dex        & km s$^{-1}$ &  km s$^{-1}$           &               &                &    (pc)        \\\hline 
 6507* & 14500$\pm$3200 & 5.5$\pm$1.0 &    20       & 112$\pm$3              & 0.68$\pm$0.05 & 15.68$\pm$0.79 & 14400$\pm$5200 \\
 8542* & 8100$\pm$4200  & 4.8$\pm$3.2 &   120       &  99$\pm$45             & 0.38$\pm$0.28 & 14.01$\pm$1.74 & 6400$\pm$5100  \\
 9227  & 9800$\pm$1800  & 4.2$\pm$0.8 &   210       &  75$\pm$13             & 0.47$\pm$0.20 & 14.21$\pm$0.78 & 7100$\pm$2800  \\
12018* & 8700$\pm$800   & 4.2$\pm$0   &    0*       &  97$\pm$6              & 0.41$\pm$0.09 & 13.52$\pm$0.39 & 5100$\pm$900   \\
13279  & 13100$\pm$1100 & 4.4$\pm$0.3 &    10       &  63$\pm$5              & 0.78$\pm$0.03 & 14.63$\pm$0.30 & 9000$\pm$1200  \\
16940  & 10800$\pm$2400 & 4.1$\pm$0.3 &   310       &  69$\pm$11             & 0.71$\pm$0.06 & 14.45$\pm$0.75 & 8200$\pm$2800  \\
24772  & 12000$\pm$1400 & 4.4$\pm$0.5 &    60       &  88$\pm$2              & 0.56$\pm$0.05 & 14.71$\pm$0.41 & 9100$\pm$1700  \\
28816  & 9700$\pm$900   & 3.6$\pm$0.5 &    70       &  96$\pm$2              & 0.50$\pm$0.08 & 13.87$\pm$0.38 & 6100$\pm$1000  \\
30971  & 11400$\pm$800  & 4.6$\pm$0.3 &    90       &  78$\pm$9              & 0.77$\pm$0.03 & 14.33$\pm$0.25 & 7800$\pm$900   \\
31183  & 14000$\pm$1600 & 4.4$\pm$0.5 &    40       &  62$\pm$2              & 0.82$\pm$0.03 & 15.22$\pm$0.41 & 11900$\pm$2200 \\
32089  & 12200$\pm$700  & 4.3$\pm$0.3 &   230       &  84$\pm$1              & 0.75$\pm$0.03 & 13.98$\pm$0.22 & 6600$\pm$900   \\\hline
\end{tabular}                                                                            
\textbf{Notes.} *: no H$_\beta$ line.
\end{table*}

\begin{table*}
\caption{Atmospheric parameters from spectroscopy of box D.}
\label{t_paramD}
\begin{tabular}{cccccc}\hline\hline
ID        &  $T_{\rm eff}$     & $\langle$RV$\rangle$  & $E(V-I)$      & $(V-Mv)_{0}$   &     d         \\
          &    (K)             & (km s$^{-1}$)         &               &                &     (pc)      \\\hline 
 7421     &  5850$\pm$40       &    76$\pm$8           & 0.35$\pm$0.03 & 13.78$\pm$0.09 & 5700$\pm$200  \\
 9011     &  6240$\pm$180      &   101$\pm$12          & 0.38$\pm$0.05 & 13.63$\pm$0.29 & 5300$\pm$700  \\
 9834     &  6210$\pm$150      &    74$\pm$3           & 0.43$\pm$0.04 & 14.20$\pm$0.24 & 6900$\pm$800  \\
11923     &  ---               &   106$\pm$8           &      ---      &       ---      &        ---    \\
19341     &  5990$\pm$200      &    88$\pm$13          & 0.38$\pm$0.05 & 13.94$\pm$0.34 & 6100$\pm$1000 \\
22319     &  5885$\pm$20       &   125$\pm$6           & 0.31$\pm$0.02 & 13.43$\pm$0.05 & 4860$\pm$100  \\
23667     &  6330$\pm$50       &   139$\pm$7           & 0.38$\pm$0.03 & 13.57$\pm$0.10 & 5200$\pm$200  \\
31274     &  5910$\pm$125      &   135$\pm$2           & 0.37$\pm$0.04 & 13.84$\pm$0.22 & 5900$\pm$600  \\
36132     &  ---               &    91$\pm$2           &      ---      &       ---      &        ---    \\
\hline                                                                                   
\end{tabular}                                                                            
\end{table*}

\section{Results of abundance analysis}
\label{s_chemres}

The chemical abundance is one of the main pillars in characterising  stellar populations, and the spectral analysis is the
most reliable technique for obtaining of the star's chemistry. The chemical pattern in the stars allows one, for example, to
distinguish which are the stars from the thick and  from thin disk \citep[e.g.,][]{mas15}, or even if a cluster (or star) has
extragalactic origin \citep[e.g.,][]{sbo15}. In this Section we present the results of our chemical analysis for the
red clump stars in order to characterise this stellar population.

   \begin{figure}
   \label{alField}
   \centering
   \includegraphics[width=\columnwidth]{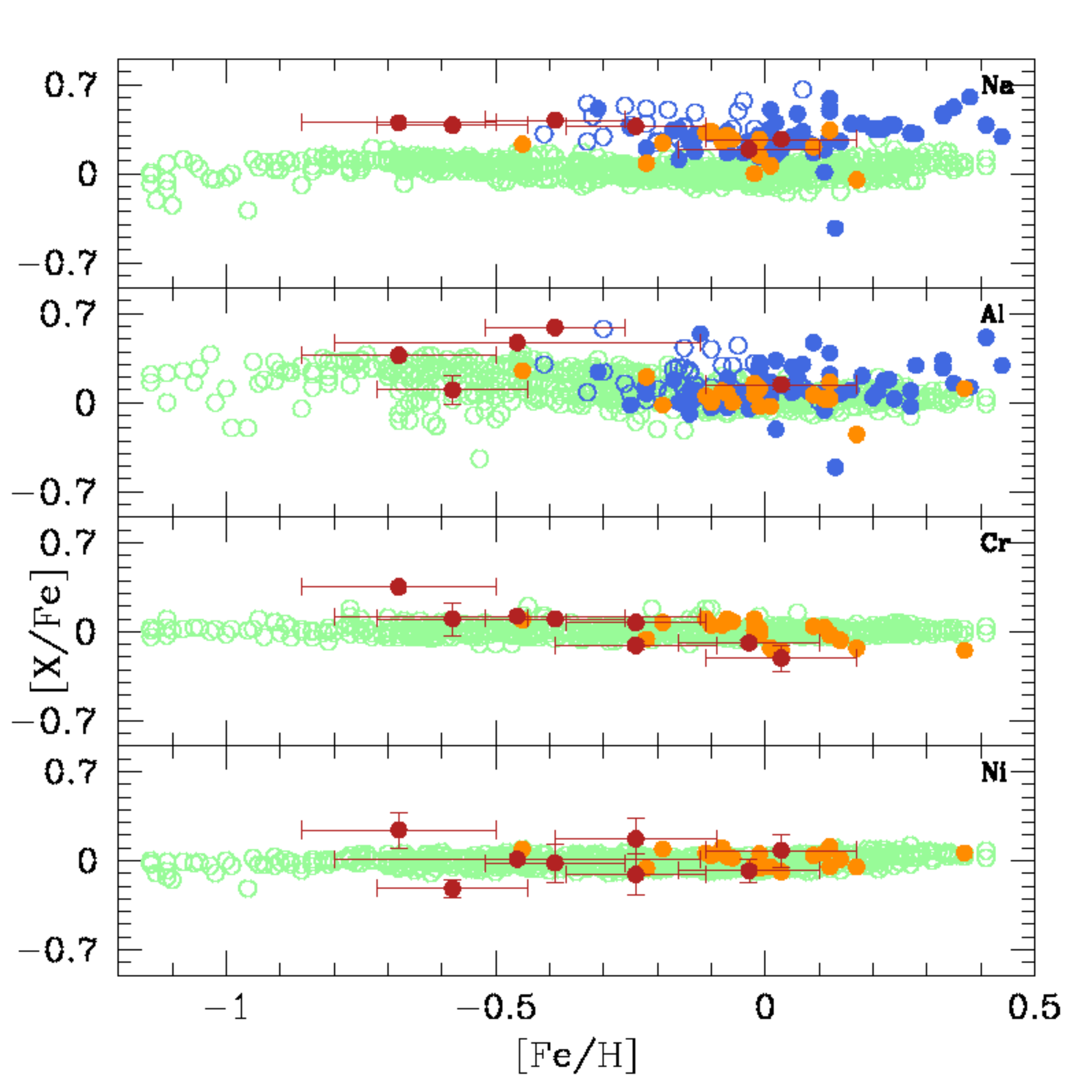}
   \caption{Abundance ratios [X/Fe] vs. [Fe/H]. Light green open circles: field dwarf of \citet{ben14}; Blue open circles: Cepheids of disk of \citet{lem13}; Blue filled circles: Cepheids of disk of \citet{gen15}; Red filled circles: our sample of red clump field stars; Orange filled circles: open clusters from literature (Tombaugh~1 of \citet{sal16}; NGC\,6192, NGC\,6404 and NGC\,6583 of \citet{mag10}; NGC\,3114 of \citet{kat13}; NGC\,2527, NGC\,2682, NGC\,2482, NGC\,2539, NGC\,2335, NGC\,2251 and NGC\,2266 of \citet{red13}; NGC\,4337 of \citet{car14b}; Trumpler 20 of \citet{car14a}; NGC\,4815 and NGC\,6705 of \citet{mag14}; Cr 110, Cr 261, NGC\,2477, NGC\,2506 and NGC\,5822 of \citet{mis15}.}%
    \end{figure}
    
   \begin{figure}
   \label{singlealphaField}
   \centering
   \includegraphics[width=\columnwidth]{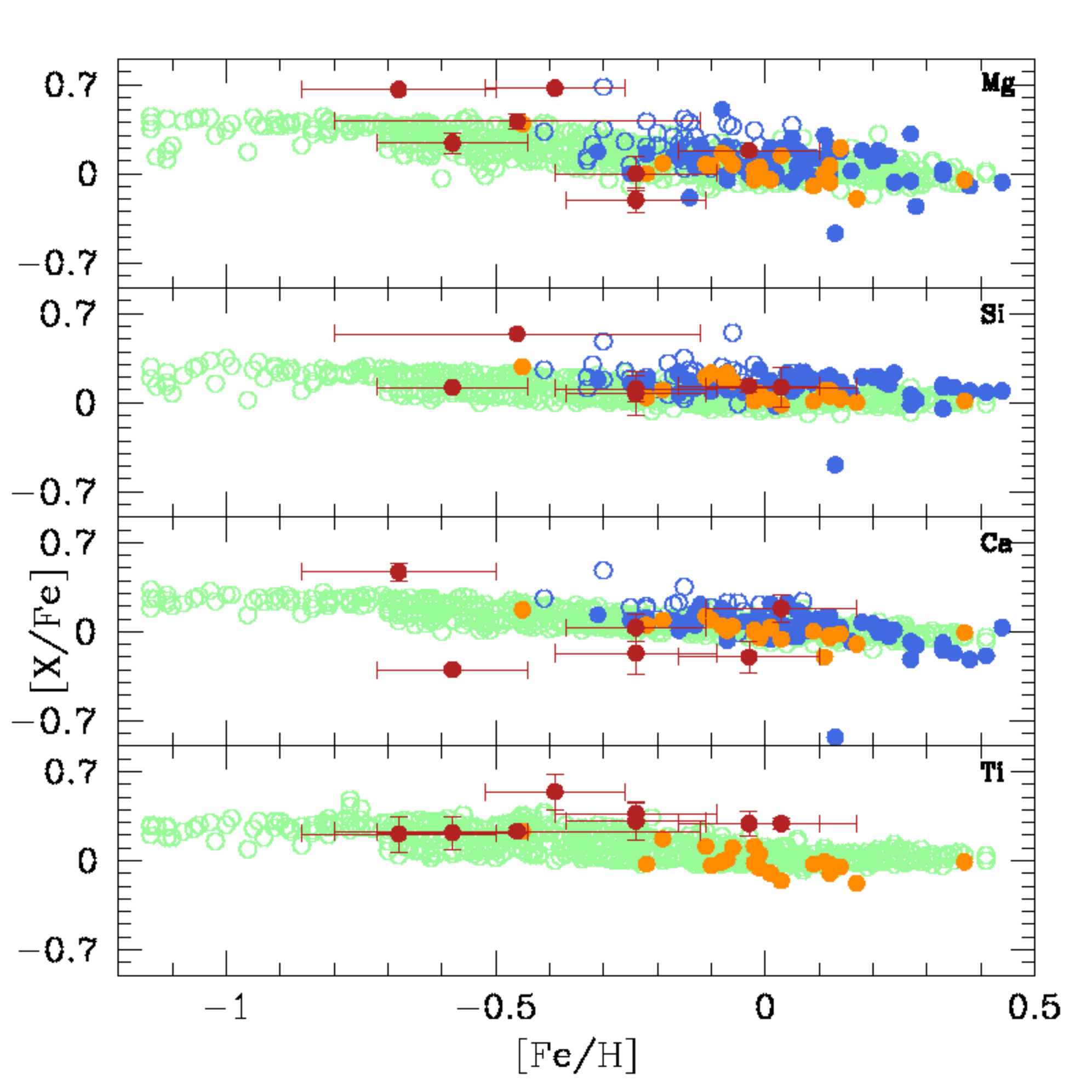}
   \caption{Abundance ratios [X/Fe] vs. [Fe/H]. Symbols have the same meaning as in Fig.~8.}%
    \end{figure}

\subsection{Metallicity and iron-peak elements}

Red clump stars have been widely used to characterise the structure of the Galaxy \citep[e.g.,][]{lee15, rom15}
mainly because they are bright enough and numerous \citep{bie14,wan15}.
In Galactic clusters, such stars are characterised by having a similar chemical abundance, unlike the field red clump stars,
which present a larger dispersion.

In Table~\ref{t_atmparamB}, we show the metallicity derived for eight red clump stars from box~B. We found that our sample of field
red clump stars covers the metallicity range of $-$0.68$\leq$[Fe/H]$\leq$0.03. Despite the wide dispersion in metallicity
of our red clump stars, some of them have similar characteristics.
Two of the targets stars with the lowest metallicity (stars $\#$13964 and $\#$28064) present high estimate for the distances (20 and 12~kpc,
respectively), although our results for the star $\#$13964 are affected by a large uncertainty in atmospheric parameters and distance due to
the low S/N of its spectrum (see Table~\ref{t_atmparamB}). Despite its higher metallicity ([Fe/H]=$-$0.24), the target $\#$31364 also
exhibits a very high distance, d$\approx$23~kpc. We also noted that these three very distant stars have slightly similar
radial velocities (73.9, 68.0 and 65.6 km/s). For the stars $\#$11029, $\#$13540, $\#$27955, and $\#$32782, our results for the metallicity
and distance indicate that they are located either near the closest edge of the outer disk or in the transition region between the outer and
inner disk (9$\leq$R$_{GC}\leq$13~kpc), where a large scatter of metallicity is found \citep{mag09}.

The Type Ia supernovae are the main sources of enrichment of the interstellar medium with Fe and iron-peak elements, like Cr and Ni.
Therefore, the chemical study of iron-peak elements is important to analyse the type Ia supernovae production rate for the formation of
the observed stellar population, being this rate, for example, a key parameter to set the Initial Mass Function (IMF). For our
sample of the red clump stars, the ranges in  abundance ratios of [X/Fe] for Cr and Ni are $-$0.09$\leq$[Cr/Fe]$\leq$+0.35 and
$-$0.22$\leq$[Ni/Fe]$\leq$+0.24. In Fig.~8 and 9 we show the abundance ratio of [X/Fe] for our sample
of red clump field star, for field dwarf from \citet{ben14}, for disk Cepheids  from \citet{lem13} and \citet{gen15}, and also for
open clusters, as described in the Figure caption.

The [Cr/Fe] and [Ni/Fe] ratios are close to solar for all our red clump targets, as observed among disk field dwarfs and open cluster
from literature in the range $-1\leq$[Fe/H]$\leq0$ (see Fig.~8). The abundance ratio of Nickel in the Milky Way, in
particular, is locked to solar value at any metallicity \citep[e.g.,][]{Sneden04}. This is usually assumed as evidence that Nickel is
synthesised in the same astrophysical sites as iron, and in a constant proportion with respect to it. However, \citet{sbo07} found that
the Sagittarius dwarf spheroidal galaxy (Sgr dSph) is depleted in Nickel by $\approx$0.3~dex in the whole range $-1\leq$[Fe/H]$\leq0$, a
behaviour that could extend even down to [Fe/H]=$-2$ \citep{sbo15}. This exotic chemical composition was, however, not observed among
metal-poor ([Fe/H]$\leq-2$) globular clusters in Fornax \citep{Letarte06}, so it is not clear if a lower [Ni/Fe] should be expected for
all dwarf galaxies at any metallicity. In any case, we find no evidence of an exotic, potentially extragalactic abundance of Nickel
in our sample.

\subsection{Na, Al and $\alpha$-elements}

Chemical abundances of the $\alpha$-elements are constantly used to reveal the history of star formation and define the structure
of the galactic disk. In the disk, stars with high abundance of $\alpha$-elements are associated with the thick disk while stars with
solar ratios are usually classified as belonging to the thin disk \citep[e.g.,][]{Bensby05}. The separation of the disk into two
components is usually interpreted as evidence that the stars of thick disk had a rapid star formation with Type-II supernovae
contributing more to the chemical enrichment of the interstellar medium than in the thin disk. In Table~\ref{abunda-NaB} we present
the abundance ratios for the alpha elements Mg, Si, Ca and Ti for the eight field red clump stars. We noted that the [X/Fe] ratios
for Ti and Si are super-solar for our sample of red clump star. For Mg, the [X/Fe] ratios shows sub-solar or solar values for two
stars ($\#$27955 and $\#$31364) while for the other red clump star we got super-solar values. And finally, the [Ca/Fe] ratio shows
super-solar values for three red clump stars and sub-solar for another three stars ($\#$11029, $\#$27955, and $\#$2806, with values $-$0.20,
$-$0.17 and $-$0.30, respectively).

On average, the stars $\#$13540, $\#$13964, $\#$26606, and $\#$32782 have a high $\alpha$-element abundances, with values of [$\alpha$/Fe] 
ratios of 0.20$\pm$0.09, 0.45$\pm$0.23, 0.39$\pm$0.16 and 0.60$\pm$0.09, respectively. Then, this stars are probably members
of the thick disk. The other red clump stars in our sample (ID~1102, 27955, 28064  and 31364) present [$\alpha$/Fe] between
0.05 and 0.10~dex, and therefore probably belong to the galactic thin disk population.

\citet{yon05}, \citet{car05}, and \citet{yon06} found that field and cluster stars in the outer galactic
disk show enhancements for the alpha-elements, [$\alpha$/Fe]$\sim$0.2, and a metallicity of approximately $\sim$-0.5
dex. However, in a more recent study of giant stars in the field of outer disk, \citet{ben11} detected thin disk
of stars with [$\alpha$/Fe]$\sim$0.0 and a lack of stars with chemical pattern of the thick disk ([$\alpha$/Fe]$\geq$0.2),
even for stars far above the Galactic plane. \citet{ben11} concluded that this lack of thick disk stars was
apparent, and was caused by the scale-length of the thick disk be significantly shorter than that of the thin disk.
Our rough estimate of the distance for the red clump stars of the box B situate many of these stars  in the outer
disk (R$_{GC}\gtrsim$13 kpc). The two disk populations (thin and thick) were detected in our sample of stars of the
outer disk. For the nearest red clump stars (1.2 $\leq$ d $\leq$ 5.3 kpc), we detected three stars with thick disk properties
(stars 1$\#$3540, $\#$26606 and $\#$32782) and two belonging to the thin disk (stars $\#$11029 and $\#$27955). The second most distant star in our
sample (star $\#$13964 with d$\sim$20 kpc) is also the star with lower metallicity ([Fe/H]=$-$0.68) and that has an
average abundance of alpha elements of [$\alpha$/Fe]=0.45$\pm$0.23. This star by its chemical pattern can be
classified as a star of the thick disk or of the Galactic disk-halo transition region. The region of Galactic disk-halo transition
is characterised by stars with $-1.20\leq$[Fe/H]$\leq$0.55, and $\alpha$-poor and $\alpha$-rich stars \citep{haw15}.
Our estimate for the distance of the star $\#$13964 (d$\sim$21 kpc), despite its significant uncertainty, may also
indicate that this star is outside the Galaxy. In this case, the star 13964 may have been lost by the Milky Way or belongs
to an extragalactic object in the vicinity of the Milky Way. The other two stars situated on the edge of outer disk 
(stars $\#$28064 and $\#$31364) have a chemical pattern typical of thin disk ([$\alpha$/Fe]$\sim$0.0). So our sample, although small,
shows that the thin disk is probably dominant in the most extreme regions of the disk, in accordance with the
conclusions of \citet{ben11}. Study of stars located in the extreme outer regions of the disk can quantify how warped and
flared is the galactic disk, as well as, how the stellar populations in these regions evolve. Furthermore, the interesting finding of
a significant amount of stars at the end of galactic disk with large estimates of distance from the galactic plane (like the stars $\#$13964,
$\#$28064, and $\#$31364) can also reveal a significant mixing between stars from the disk and the halo caused by the warped and flared disk. But
what this interaction between the warped and flared disk and the halo implies for  the evolution of stellar populations in this
extreme region of the Galaxy? In this region, do we have a stellar population predominantly $\alpha$-rich or $\alpha$-low? And what is
the metallicity range? A major difficulty for a reliable study of the structure of extreme outer region of the disk is an estimated
distance that often comes with large uncertainties (as our estimate for the star 13964). Incoming data
from Gaia mission will surely enable more solid distance estimates for many stars and put us in a better position to answer to these questions.

The production of the Na in the stellar interior is performed during the hydrostatic carbon burning in massive stars \citep{woo95},
and also is affected by NeNa cycle in the H-burning envelope in intermediate-mass and massive stars \citep{den90}. In giant stars,
the chemical abundance of Na is important to investigate the mixing processes occurring in the stellar interior, like the first
dredge-up, {\it thermohaline} instability, and rotation-induced mixing \citep{cha10}.

The effects of the non-local thermodynamic equilibrium (NLTE) are considerable in the abundance of Na \citep{gra99,lin11} and should
be taken into account. In our analysis, we used the corrections of \citet{gra99} to
estimate such effects. Table~\ref{abunda-NaB} show sthe [Na/Fe] ratios for six red clump stars. The abundance ratio [Na/Fe] for
our field red clump stars presents an overabundance that goes from 0.19 to 0.42. In Fig.~\ref{alField} we see that the [Na/Fe] ratio
for our sample are overabundant when compared to  disk  dwarfs  from \citet{ben14}. The overabundance of [Na/Fe] ratio in giant
stars with respect dwarf stars indicates the importance of the chemical mixing phenomena occurring in the stellar interior during
the giant phase \citep{pas04}.

Al is mainly produced during the hydrostatic carbon and neon burning in massive stars \citep{woo95}, and can be affected by MgAl
cycle in H-burning layers at high temperatures \citep{arn95}. We observed an overabundance of Al with respect to Fe for our red clump
sample, with a range of 0.10$\leq$[Al/Fe]$\leq$0.59 (see Table~\ref{abunda-NaB}). The stars that present the highest values for the
[Al/Fe] ratio ($\#$13964, $\#$26606, and $\#$32782) are also the stars that have a high overabundance of $\alpha$-elements. 

While $\alpha$-elements are overabundant in stellar populations characterised by a fast star formation history such as the Galactic
thick disk, the stars in dwarf galaxies are usually $\alpha$-depleted. This is because the slow star formation rate in these
low-density environments prevents the yields of Type-II supernovae from dominating the pollution of the interstellar medium. In
fact, the average $\alpha$-elements abundance of Sgr dSph stars in the range $-1\leq$[Fe/H]$\leq0$ is [$\alpha$/Fe]$\approx-0.2$
\citep{sbo07}. Similar sub-solar $\alpha$-abundances were found by \citet{sbo05} in three field stars, and they were the basis of
their claim of an extragalactic origin for their targets. Na and Al also are depleted in intermediate-metallicity Sgr
dSph stars by $\approx-0.3$ and $-0.5$~dex, likely because the astrophysical sites of their synthesis, massive and intermediate-mass
stars, is the same where most of the $\alpha$-elements are produced. In this context, we note that none of our box~B stars show
abundances that differ from the Galactic trend. The $\alpha$-elements, Na, and Al abundances, along with Ni that we discussed
in the previous Section, are either close to solar, or super-solar for the most metal-poor targets, in full agreement with the
general trend of Galactic thin and thick disk stars. We therefore conclude that there is no evidence of an extragalactic origin for
any of the studied object.

In the discussion section we put the red clump stars together with the other stars of our sample in the context of the 
structure of the third Galactic quadrant and try to make connections between the different populations so far analysed .

\begin{figure}
\label{f_reddist}
\centering
\includegraphics[width=\columnwidth]{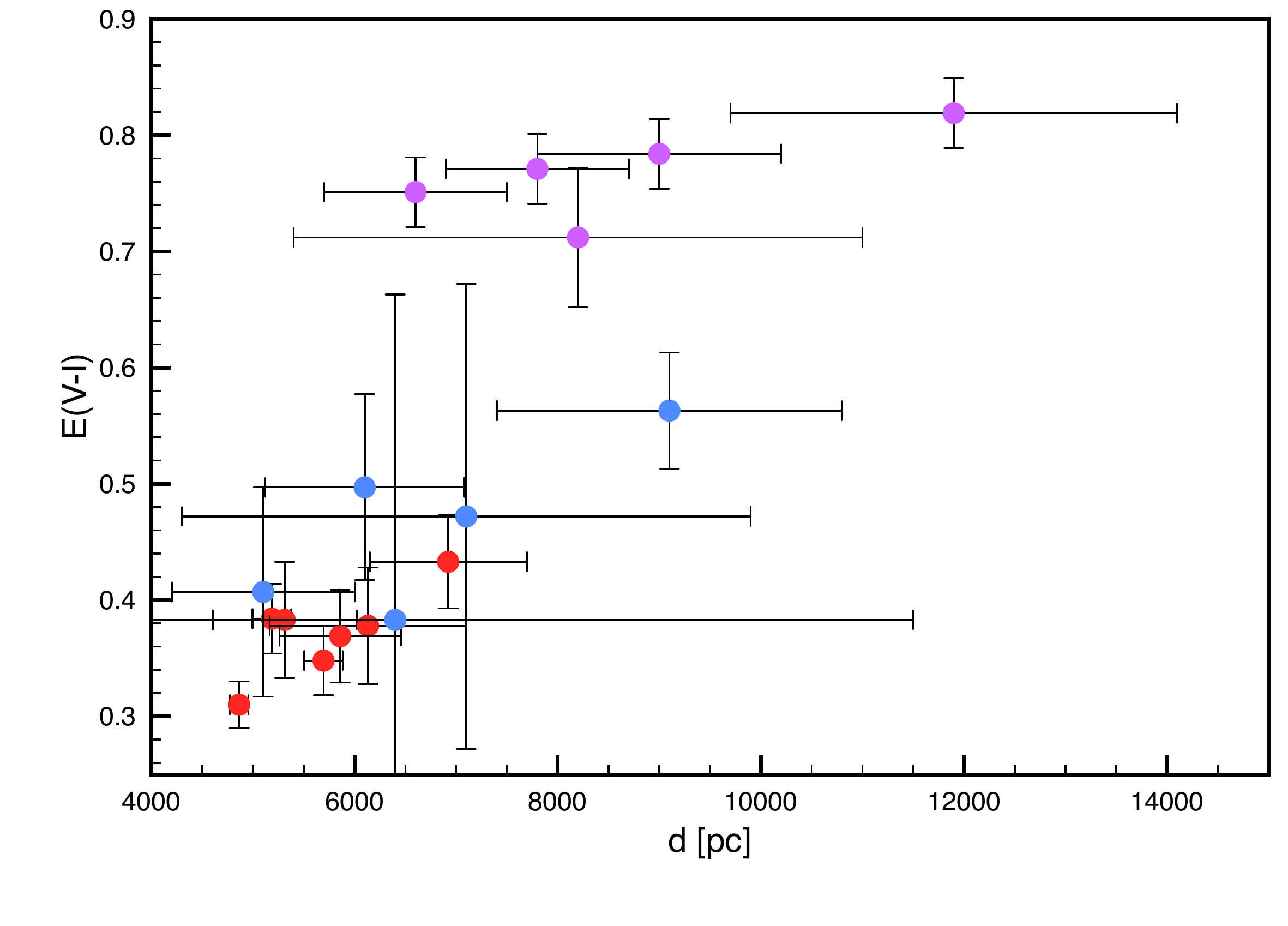}
\caption{Spectroscopic results of reddening as a function of distance for the targets in Box~C and D. The blue, violet, and red dots
indicate the C1, C2, and D groups, respectively.}
\end{figure}

\section{Discussion and conclusions}

In the following we are going to discuss the results of our photometric and spectroscopic analysis and attempt to draw a coherent
scenario out of them. \\

The targets in box~D (see Fig.~1) belong to a thick, faint MS in the background of the cluster Tombaugh~1. The
observed objects (see Table~\ref{t_paramD}) are all F8-G2 spectral type stars with similar reddening, spanning a narrow range in
distance (5--7~kpc, 5.7~kpc on average). They trace a tight sequence in the CMD. MS stars with these spectral types must be younger
than $\approx$9~Gyr, and the MS in the CMD continues blue-ward to even higher temperatures (group in box C, see below). Hence, this
MS traces an intermediate-age stellar population, and it cannot be associated to the Galactic halo or thick disk. The distance
spread is most likely physical and not only a product of measurement errors, because the stars follow a clear reddening-distance
relation, as shown in Fig.~10. Two stars slightly depart from this relation, probably due to differential reddening in
the field of view. The shape of this sequence and its width are comparable with the ones found in the background of NGC~2354
\citep{car16}, or in the direction of the Canis Major over density \citep{car08}, where these sequences are ascribed to the
warped old thin disk, that the line of sight intersects, thus producing the effect of probing a structure confined in distance.

The weighted average RV is $\sim$107~km~s$^{-1}$, much higher than the expectation of a simple Galactic rotation model
($\approx$60~km~s$^{-1}$) such as that presented by \citet{mon14}. However, the model could easily fail at these large distances
from the Galactic center, and far away from the formal Galactic plane (at latitude 0$^o$ deg). The line-of-sight velocity at this
Galactic longitude mostly reflects the Galactocentric $U$ component. The RV dispersion, after quadratic subtraction of the mean
observational error, is $\sigma_{RV}=23$~km~s$^{-1}$. This is very high, because $\sigma_U\approx10$~km~s$^{-1}$ for the thin disk
at the solar position, and the dispersion is expected to exponentially decline outwards. Hence, this intermediate-age and distant
population presents peculiar kinematical properties. A possible explanation for  this peculiarity is that this population belongs to
the Galactic warp \citep{mom06}, and it is also flared  \citep[see][and references therein]{car15}. In this scenario the kinematics
is not easily predictable (see Xu et al. 2015), since the outer disk exhibits several rings and waves, which alter the expected
kinematics.

\noindent
We now continue discussing  stars in box~C. These stars form a sequence which lies along the prolongation of the thick MS we just
described. In this case they can be interpreted as blue straggler stars of this intermediate-age population. Their color and
magnitude can also be compatible with them being thick disk or halo foreground hot sdB’s, although we do not expect so many
stars of this type in such a limited volume \citep{car15}. The spectroscopic data we have analysed in this work help us to
understand better the nature of these stars.

\begin{figure}
\label{f_cmdCD}
\centering
\includegraphics[width=\columnwidth]{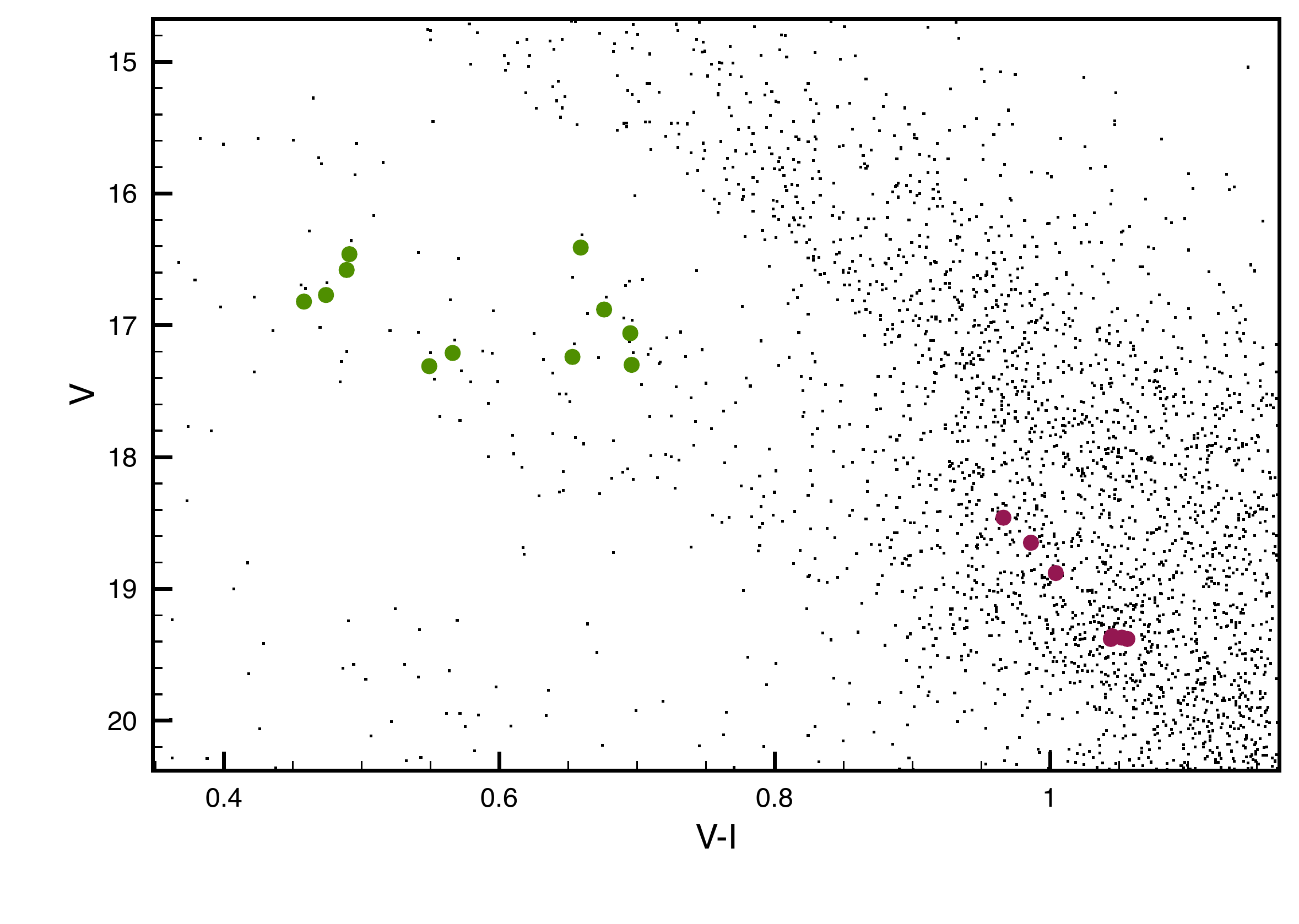}
\caption{CMD of the field, with box~C and box~D targets marked with green and red symbols, respectively.}
\end{figure}

Reading through Table~\ref{t_paramC} one can infer that these C group stars are early type stars, confirming earlier findings
\citep[][and references therein]{car16} based on photometry only. Therefore we remark that the {\it blue plumes} routinely found in many
different lines of sight in the third quadrant of the Milky Way \citep{moi06} are indeed sequences of young stars. Only one of the
observed targets is likely a sdB star, as commented in Sect.~\ref{s_CDparams} and it will be excluded from further discussion. For
this specific line of sight, a quick glance at the CMDs in Fig.~1, 4, and 11 reveals that the stars selected for, and observed with,
spectroscopy are clearly separated in the CMD, where there are five stars with $(V-I)<0.57$ (hereafter C1 group) and five object at
$(V-I)>0.65$ (hereafter C2). The two groups seem to trace two separated sequences. The same dichotomy is found in the spectroscopic
results. All C2 objects have $T_{eff}>10\,000~K$, while four of the five C1 stars are cooler than 10\,000~K. Thus, C2 stars are on
average hotter, despite they are redder in the CMD, and in fact they exhibit a much larger reddening ($\overline{E(V-I)}=0.46$ and
0.77, for C1 and C2 groups, respectively). 

The two groups also show a different behavior in the distance-reddening relation shown in Fig.~10. In fact, C1 stars are
compatible with the distance-reddening relationship defined by box~D targets, and they are distributed in a distance range that largely
overlaps that of box~D stars, with a mean value of $d=6.7$~kpc. C2 stars, on the other hand, are found at nearly constant reddening, and on
average at a larger distance than box~D and C1 stars ($d=8.2$~kpc on average). In addition, the kinematics of the two C-groups also seems
different. C1 stars are are confined in a narrow range of RVs between 75 and 100~km~s$^{-1}$, with a mean value of $91$~km~s$^{-1}$, similar
to that found in box~D. The mean RV of C2 stars, on the other hand, is 70~km~s$^{-1}$, matching within few km~s$^{-1}$ the expectations of a
simple Galactic rotation model at $d=8$~kpc. Their RV dispersion is also low, 8.2~km~s$^{-1}$, as expected for a thin disk population.

The observed differences between C1 and C2 stars could be partially due to differential reddening. In Fig.~12 we show the
position of the targets in the \citet{Schlegel98} reddening map. There is clearly a variation of reddening in the field, and the C1 (C2)
stars tend to be found far from (close to) a local reddening maximum. However, a further inspection suggests that this cannot fully explain
the observed dichotomy. First, there is a certain degree of mixing in the spatial distribution, with a C2 target found in a low-reddening
area and the object closest to the reddening maximum being a C1 star. In addition, the reddening variations are small in the
\citet{Schlegel98} map, where it varies by no more than 0.1 mag in the field under investigation. This is only one third of the difference
between the average reddening of C1 and C2 groups. The variation would be even reduced if corrections to the maps, such as that proposed by
\citet{Bonifacio00}, were applied. The differential reddening also cannot explain the different kinematics of the two groups, nor the
distance-reddening relation observed for C1 and D stars. Moreover, spatial reddening variations alone would cause that more reddened stars
are closer, contrary to what observed, because the targets in each box were selected photometrically at approximately the same magnitude. We
conclude that differential reddening may play a role, but it cannot alone explain the observed differences between C1 and C2 groups.

\begin{figure}
\label{f_Schlegel}
\centering
\includegraphics[width=\columnwidth]{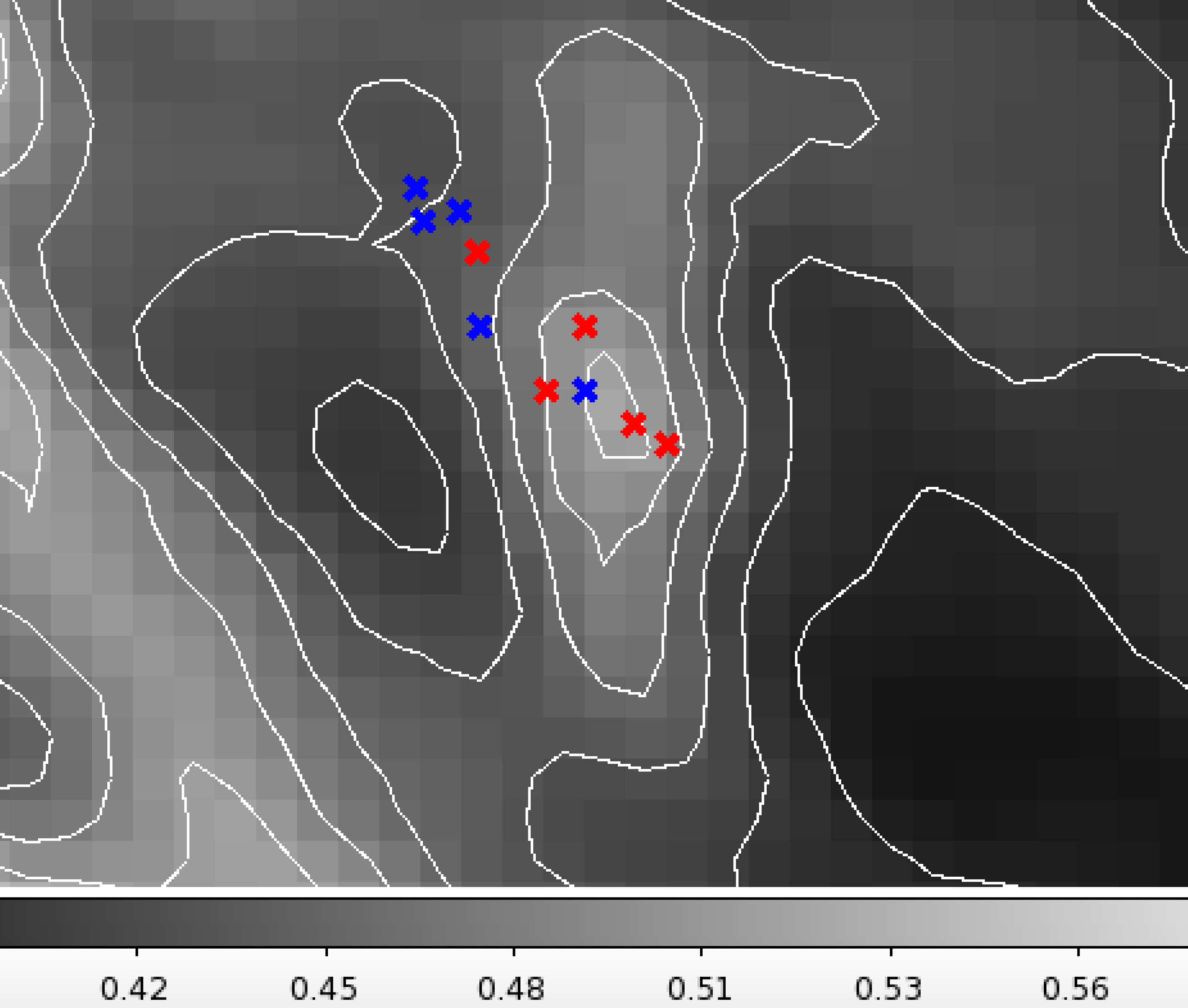}
\caption{A reddening map from Schlegel et al (1998) in the region of Tombaugh 1. Blue and red crosses indicate stars from C1 and C2 sub-groups, respectively.}
\end{figure}

We have to take these differences with a lot of care, since they are derived from a handful of stars. From the photometric analysis
in fact we see a continuum of reddening and distance properties, where these two groups, C1 and C2, are representing the extremes of
these young population, as seen from the CMD in Fig.~11. This discrepancy between the photometric and spectroscopic
distribution of reddening and distances in Box~C can offer two alternative interpretations: $i$) the distribution is broad and
continuum, as indicated by the photometric results, and the spectroscopic dichotomy is only an spurious effect due to the random
selection a small sample of targets, or due to selection biases; $ii$) The underlying distribution is bimodal, as the spectroscopic
results suggest, but its nature is lost in the photometric results due to large uncertainties, which blend the two groups into a
wider single peak. To investigate this issue in more details, we compare in Fig.~13 the distribution of $E(V-I)$
obtained with the spectroscopic and photometric method, for the same ten targets of the spectroscopic study. The histogram shows
that the bimodal distribution found with the first method is completely lost when the photometrically-derived reddening are studied.
This results suggests that, if the C1 and C2 groups represent two distinct stellar populations with different distances and
reddening in the same field of view, their presence would have likely been missed in the photometric results. A Kolmogorov-Smirnov
test indicates that the probability that the bimodal spectroscopic results are randomly drawn from the photometric broad distribution
is only 8\%. Hence, the hypothesis that this multi-modality is entirely due to the random selection of a small sample of stars cannot
be dismissed, but it is very improbable. However, while these tests tend to exclude the hypothesis $i$) above, they are insufficient
to claim the existence of two stellar sub-structures at different distances in this field, both because of the too small
spectroscopic sample, and because selection effects unaccounted for in the analysis could have led to a observed sample unevenly
distributed in the CMD, thus giving the wrong impression of a bimodal distribution. A more extensive spectroscopic study of a larger
sample of stars is required to fix this issue.

This young population is the very same that we found in several other direction in the third Galactic quadrant. It is confined in
distance, being at heliocentric distances in the range 6 to 9~kpc. Within the uncertainties involved, these stars are most probably
tracing a portion of the outer, or Norma-Cygnus, spiral arm. This arm is located well below the formal Galactic plane (at b =0$^o$),
because of the warp, and the line of sight to Tombaugh~1 intersects it, in close similarity to the line of sight to the old open cluster Auner~1
(Carraro et al. 2007). Being the Norma-Cygnus arm the outermost arm known for the
Milky Way, it is not unexpected to find stars at so very different distances, because outermost arms are typically wider than
inner disk arms, whose width is typically about 1~kpc. Interestingly, the line of sight to Tombaugh~1 does not contain young stars closer to
the Sun (at 1.5$-$2.5 kpc), that we would expect from the crossing of the Perseus arm, which would be located at about 2~kpc from the Sun
\citep{church09}. The fact that
we miss the Perseus arm means either that the warping of the disk starts to be significant beyond 2-3~kpc, or that the Perseus arm is
not important in the third Galactic quadrant.

\begin{figure}
\label{f_histoEVI_C}
\centering
\includegraphics[width=\columnwidth]{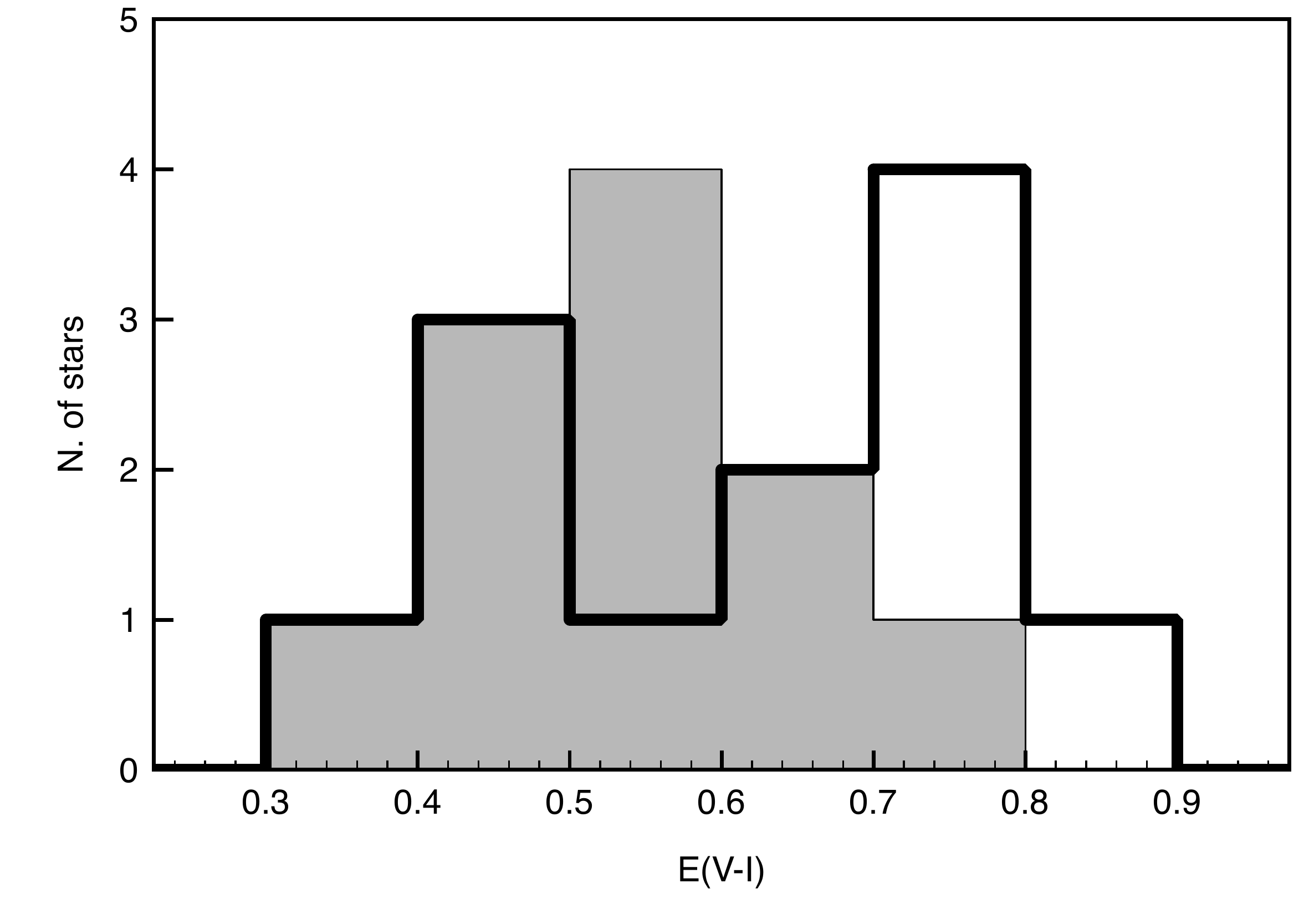}
\caption{Histogram of the spectroscopic (thick black line) and photometric (shaded grey area)reddening distribution of the box~C stars.}
\end{figure}

Finally, we comment on the stars located in the Box~B. This box samples red giant stars. According to our results (see
Tables~\ref{t_atmparamB} and \ref{abunda-NaB}), these stars, which seem to somehow group together in the CMD, are located at
different distances, and have different metallicities. Thus, the stars of box B have no close relation to the populations of the
box~C and D. As discussed in Sect.~\ref{sec:intro}, the presence of a background MS and a {\it blue plume} of hot young stars in the
third Galactic quadrant have been interpreted as evidence of the recent accretion of the Canis Major satellite. We argued in
Sect.~\ref{s_chemres} that we find no evidence of an extragalactic origin for any of our red clump targets. Unfortunately, the lack
of a kinematical or spatial link of box~B stars with the fainter box~C and D groups, prevents us to extend the result to these
features. However, red clump stars should be present in a complex stellar population as that postulated for the Canis Major
satellite, and our sample included targets in a wide range of distance. Our results therefore favours the scenario where the peculiar
features observed in the CMD are due to a complex mix of Galactic populations rather than the imprint of an extragalactic system.

We can divide the stars of box B in two field populations. The first composed by older stars, with ages of 8~Myr and 10~Gyr,
belonging to the thick and thin disk (with slight majority for the thick disk) and a distance between 1.2$\leq$d$\leq$5.3~kpc. The
second population composed by young ($\lesssim$8~Myr) background field stars, with high values for distances (d$>$12~kpc) and slightly
similar radial velocities, with mean RV of 69~km~s$^{-1}$. It is mportant to note that the distances to these background stars of the box~B
have large  uncertainties with average of $\sim$9~kpc being the important result for these stars that they are background
stars relating the other stars of B box. It is worth mentioning that the detection of stars with large estimates of the distances, and
consequently with large distance from the galactic plane, as the stars $\#$13964, $\#$28064, and $\#$31364 in box B, and also apparently young
when compared to thick disk stars and halo, is not expected, and can reveal an interesting and complex galaxy evolution occurring in
the pherifery of the warped and flared outer disk with a probable interaction with the Galactic halo.
A contributing to this discussion comes from recent Galactic disk simulations reveal that younger populations have
increasingly larger scale-lengths at larger distances from the galactic center \citet{min15}. Such featured reveals the need for
further analysis with a large sample of stars in this extreme region of the disk.
The Gaia-ESO survey can help us to answer these questions better, since this survey will enable a reliable determination of the distances of large 
samples of  these stars.

\acknowledgments
We express our deepest gratitude to Edgardo Costa, who acquired part of the data used for this work.
Extensive use was made of the WEBDA database maintained by E. Paunzen at the University of Vienna, Austria (http://www.univie.ac.at/webda). 
J.V.S.S. acknowledges the support provided by CNPq/Brazil Science
without Borders program (project No. 249122/2013-8). C.M.B. acknowledges support by the Fondo Nacional de Investigaci\'on Cient\'ifica y
Tecnol\'ogica (Fondecyt), project No. 1150060.

\vspace{5mm}







\begin{thebibliography}{}
\bibitem[Alonso et al.(1999)]{alo99} Alonso, A., Arribas, S., Mart\'inez-Roger, C., 1999, A\&AS, 140, 261
\bibitem[Arnould \& Mowlavi(1995)]{arn95} Arnould, M., \& Mowlavi, N.\ 1995, Liege International Astrophysical Colloquia, 32, 17
\bibitem[Asplund et al.(2009)]{asp09} Asplund, M., Grevesse, N., Sauval, A.~J., \& Scott, P. 2009, ARA\&A, 47, 481
\bibitem[Bellazzini et al.(2004)]{bel04} Bellazzini, M., Ibata, R., Monaco, L., Martin, N., Irwin, M.~J. \& Lewis, G.~F. 2004, MNRAS, 354, 1263
\bibitem[Bensby et al.(2005)]{Bensby05} Bensby, T., Feltzing, S., Lundstr\"{o}m, I., \& Ilyin, I. 2005, \aap, 433, 185
\bibitem[Bensby et al.(2011)]{ben11} Bensby, T., Alves-Brito, A., Oey, M.~S., Yong, D., \& Mel{\'e}ndez, J.\ 2011, \apjl, 735, L46
\bibitem[Bensby et al.(2014)]{ben14} Bensby, T., Feltzing, S., \& Oey, M.~S.\ 2014, \aap, 562, A71
\bibitem[Bergeron et al.(1992)]{Bergeron92} Bergeron, P., Saffer, R.~A., \& Liebert, J. 1992, ApJ, 394, 228
\bibitem[Bertelli et al.(2008)]{bert08} Bertelli, G., Girardi, L., Marigo, P., Nasi, E. 2008, A\&A, 484, 815
\bibitem[Bertelli et al.(2009)]{bert09} Bertelli, G., Nasi, E., Girardi, L., Marigo, P. 2009, A\&A, 508, 355
\bibitem[Bessell et al.(1998)]{bes98} Bessell, M.~S., Castelli, F., \& Plez, B.\ 1998, \aap, 333, 231
\bibitem[Bienaym\'e et al.(2014)]{bie14} Bienaym\'e, O., Famaey, B., Siebert, A., et al. 2014, A\&A, 571, 92
\bibitem[Blackwell et al.(1986)]{bla86} Blackwell, D.~E., Booth, A.~J., Menon, S.~L.~R., \& Petford, A.~D. 1986, MNRAS, 220, 289
\bibitem[Bonifacio et al.(2000)]{Bonifacio00} Bonifacio, P., Monai, S., \& Beers, T.~C. 2000, AJ, 120, 2065
\bibitem[Carney et al.(2005)]{car05} Carney, B.~W., Yong, D., Teixera de Almeida, M.~L., \& Seitzer, P.\ 2005, \aj, 130, 1111
\bibitem[Carraro et al.(2005)]{carr05} Carraro, G., V{\'a}zquez, R.~A., Moitinho, A., \& Baume, G.\ 2005, \apjl, 630, L153
\bibitem[Carraro et al.(2007)]{car07} Carraro, G., Moitinho, A., Zoccali, M., Vazquez, R., Baume, G., 2007, \aj, 133, 1058
\bibitem[Carraro et al.(2008)]{car08} Carraro, G., Moitinho, A., \& V{\'a}zquez, R.~A.\ 2008, \mnras, 385, 1597
\bibitem[Carraro et al.(2010)]{car10} Carraro, G., V\'azquez, R.~A., Costa, E., Perren, G., \& Moitinho, A. 2010, ApJ, 718, 683
\bibitem[Carraro et al.(2014)]{car14a} Carraro, G., Monaco, L., Villanova, S., 2014a,  A\&A, 568, 86
\bibitem[Carraro et al.(2014)]{car14b} Carraro, G., Villanova, S., Monaco, L., Beccari, G., Ahumada, J.,  Boffin, H., 2014b, A\&A, 562, 39
\bibitem[Carraro et al.(2015)]{car15} Carraro, G., V{\'a}zquez, R.~A., Costa, E., Ahumada, J.~A., \& Giorgi, E.~E.\ 2015, \aj, 149, 12
\bibitem[Carraro et al.(2016)]{car16} Carraro, G., Seleznev, A.~F., Baume, G., \& Turner, D.~G.\ 2016, \mnras, 455, 4031
\bibitem[Carretta et al.(2004)]{car04} Carretta, E., Bragaglia, A., Gratton, R.~G., \& Tosi, M. 2004, A\&A, 422, 951
\bibitem[Carretta et al.(2007)]{car07} Carretta, E., Bragaglia, A., \& Gratton, R.~G. 2007, A\&A, 473, 129
\bibitem[Castro et al.(1997)]{cas97} Castro, S., Rich, R.~M., Grenon, M., Barbuy, B., \& McCarthy, J. K. 1997, AJ, 114, 376
\bibitem[Charbonnel \& Lagarde(2010)]{cha10} Charbonnel, C., \& Lagarde, N. 2010, A\&A, 522, A10
\bibitem[Churchwell et al.(2009)]{church09} Churchwell, E., Babler, B.~L., Meade, M.~R., Whitney, B.~A., Benjamin, R., et al., 2009, PASP, 121, 213 
\bibitem[Coelho et al.~(2005)]{Coelho05} Coelho, P., Barbuy, B., Mel\'endez, J., Schiavon, R.~P., \& Castilho, B.~V. 2005, A\&A, 443, 735
\bibitem[Denisenkov \& Denisenkova(1990)]{den90} Denisenkov, P.~A., \& Denisenkova, S.~N. 1990, SvAL, 16, 275
\bibitem[Depagne et al.~(2002)]{dep02} Depagne, E., Hill, V., Spite, M., et al. 2002, A\&A, 390, 187
\bibitem[Drake \& Smith(1991)]{dra91} Drake, J.~J., \& Smith, G. 1991, MNRAS, 250, 89
\bibitem[Dressler et al.(2006)]{dre06} Dressler, A., Hare, T., Bigelow, B.~C., \& Osip, D.~J. 2006, Proc. SPIE, 6269, 62690
\bibitem[Edvardsson et al.(1993)]{edv93} Edvardsson, B., Andersen, J., Gustafsson, B., Lambert, D.~L., Nissen, P.~E., \& Tomkin, J. 1993, A\&A, 275, 101
\bibitem[Fuhrmann et al.(1994)]{Fuhrmann94} Fuhrmann, K., Axer, M., \& Gehren, T. 1994, A\&A, 285, 585
\bibitem[Genovali et al.(2015)]{gen15} Genovali, K., Lemasle, B., da Silva, R., et al.\ 2015, \aap, 580, A17
\bibitem[Gratton \& Sneden(1988)]{gra88} Gratton, R.~G., \& Sneden, C. 1988, A\&A, 204, 193
\bibitem[Gratton et al.(1999)]{gra99} Gratton, R.~G., Carretta, E., Eriksson, K., \& Gustafsson, B. 1999, A\&A, 350, 955
\bibitem[Grevesse \& Sauval(1998)]{gre98} Grevesse, N., \& Sauval, A. J. 1998, SSRv, 85, 161
\bibitem[Hawkins et al.(2015)]{haw15} Hawkins, K., Jofr{\'e}, P., Masseron, T., \& Gilmore, G.\ 2015, \mnras, 453, 758
\bibitem[Horne(1986)]{hor86} Horne, K. 1986, PASP, 98, 609
\bibitem[Katime Santrich et al.(2013)]{kat13} Katime Santrich, O.~J., Pereira, C.~B., \& de Castro, D.~B.\ 2013, \aj, 146, 39
\bibitem[Keeping(1962)]{kee62} Keeping, E.~S. 1962, Introduction to Statistical Inference (London: Van Nostrand)
\bibitem[Kuijken \& Gilmore(1989)]{Kuijken89} Kuijken, K., \& Gilmore, G. 1989, MNRAS, 239, 605
\bibitem[{{Kurucz}(1993)}]{kur93} Kurucz, R. 1993, ATLAS9 Stellar Atmosphere Programs and 2 km/s grid.~Kurucz
  CD-ROM No.~13.~Cambridge, Mass.: Smithsonian Astrophysical Observatory, 1993, 13
\bibitem[Lambert et al.(1996)]{lam96} Lambert, D.~L., Heath, J.~E., Lemke, M., \& Drake, J. 1996, ApJS, 103, 183
\bibitem[Landolt(1992)]{lan92} Landolt, A. U. 1992, AJ, 104, 372
\bibitem[Lee et al.(2015)]{lee15} Lee, Y.-W., Joo, S.-J., \& Chung, C.\ 2015, \mnras, 453, 3906
\bibitem[Lemasle et al.(2013)]{lem13} Lemasle, B., Fran\c{c}ois, P., Genovali, K., et al. 2013, A\&A, 558, 31
\bibitem[Letarte et al.(2006)]{Letarte06} Letarte, B., Hill, V., Jablonka, P., Tolstoy, E., Fran\c{c}ois, P., \& Meylan, G. 2006, A\&A, 453, 547
\bibitem[Lind et al.(2011)]{lin11} Lind, K., Asplund, M., Barklem, P.~S., \& Belyaev, A.~K. 2011, A\&A, 528, A103
\bibitem[Magrini et al.(2009)]{mag09} Magrini, L., Sestito, P., Randich, S., \& Galli, D. 2009, A\&A, 494, 95
\bibitem[Magrini et al.(2010)]{mag10} Magrini, L., Randich, S., Zoccali, M., Jilkov\'a, L., Carraro, G., et al., 2010, A\&A, 523, 11
\bibitem[Magrini et al.(2014)]{mag14} Magrini, L., Randich, S., Romano, D., Friel, E.D., et al., 2014, A\&A, 563, 44
\bibitem[Majaess et al.(2013)]{Majaess13} Majaess, D., Carraro, G., Moni Bidin, C., et al. 2013, A\&A, 560, A22
\bibitem[Martin et al.(2002)]{mar02} Martin, W.~C., Fuhr, J.~R., Kelleher, D.~E., et al. 2002, NIST Atomic Spectra Database (Version 2.0; Gaithersburg, MD: NIST)
\bibitem[Martin et al.(2004)]{mar04} Martin, N.~F., Ibata, R.~A., Bellazzini, M., et al.\ 2004, \mnras, 348, 12
\bibitem[Masseron \& Gilmore(2015)]{mas15} Masseron, T., \& Gilmore, G.\ 2015, \mnras, 453, 1855
\bibitem[McWilliam and Rich(1994)]{mcw94} McWilliam, A., \& Rich, R. M. 1994, ApJS, 91, 749
\bibitem[Minchev et al.(2015)]{min15} Minchev, I., Martig, M., Streich, D., et al.\ 2015, \apjl, 804, L9
\bibitem[Mishenina et al.(2015)]{mis15} Mishenina, T., Pignatari, M., Carraro, G., et al. 2015, MNRAS, 446, 3651
\bibitem[Moitinho(2001)]{moitinho01}, Moitinho, A., 2001, A\&A, 370, 436
\bibitem[Moitinho et al.(2006)]{moi06} Moitinho, A., V\'azquez, R.~A.,Carraro, G., Baume, G., Giorgi, E.~E.,\& Lyra, W. 2006, MNRAS, 368, 77
\bibitem[Momany et al.(2006)]{mom06} Momany, Y., Zaggia, S., Gilmore, G., et al.\ 2006, \aap, 451, 515
\bibitem[Moni Bidin et al.(2010)]{moni10} Moni Bidin, C., de la Fuente Marcos, R., de la Fuente Marcos, C., \& Carraro, G. 2010, A\&A, 510, A44
\bibitem[Moni Bidin et al.(2011)]{mon11} Moni Bidin, C., Villanova, S., Piotto, G., \& Momany, Y. 2011b, A\&A, 528, A127
\bibitem[Moni Bidin et al.(2012a)]{mon12a} Moni Bidin, C., Carraro, G., \& M\'endez, R.A. 2012, ApJ, 747, 101
\bibitem[Moni Bidin et al.(2012b)]{moni12} Moni Bidin, C., Villanova, S., Piotto, G., Moehler, S.,Cassisi, S., \& Momany, Y. 2012, A\&A, 547, A109
\bibitem[Moni Bidin et al.(2014)]{mon14} Moni Bidin, C., Majaess, D., Bonatto, C., et al. 2014, A\&A, 561, 119
\bibitem[Morse et al.(1991)]{mor91} Morse, J.A., Mathieu, R.D., \& Levine, S.E. 1991, AJ, 101, 1495
\bibitem[Munari et al.(2005)]{mun05} Munari, U., Sordo, R., Castelli, F., \& Zwitter, T. 2005, A\&A, 442, 1127
\bibitem[Napiwotzki et al.(1999)]{Napiwotzki99} Napiwotzki, R., Green, P.~J., \& Saffer, R.~A. 1999, ApJ, 517, 399
\bibitem[Nieva \& Przybilla(2007)]{Nieva07} Nieva, M.~F., \& Przybilla, N. 2007, A\&A, 467, 295
\bibitem[Pasquini et al.(2004)]{pas04} Pasquini, L., Randich, S., Zoccali, M., et al.\ 2004, \aap, 424, 951
\bibitem[Patat \& Carraro(2001)]{pat01} Patat, F., \& Carraro, G. 2001, MNRAS, 325, 1591
\bibitem[Pereira et al.(2011)]{per11} Pereira, C.~B., Sales Silva, J.~V., Chavero, C., Roig, F., \& Jilinski, E.\ 2011, \aap, 533, A51
\bibitem[Preston \& Sneden(2001)]{pre01} Preston, G.~W., \& Sneden, C.\ 2001, \aj, 122, 1545
\bibitem[Reddy et al.(1999)]{red99} Reddy, B.~E., Bakker, E.~J., \& Hrivnak, B.~J. 1999, ApJ, 524, 831
\bibitem[Reddy et al.(2003)]{red03} Reddy, B.~E., Tomkin, J., Lambert, D.~L., \& Allende Prieto, C. 2003, MNRAS, 340, 304
\bibitem[Reddy et al.(2013)]{red13} Reddy, A.~B.~S., Giridhar, S., \& Lambert, D.~L. 2013, MNRAS, 431, 3338
\bibitem[Romero-G{\'o}mez et al.(2015)]{rom15} Romero-G{\'o}mez, M., Figueras, F., Antoja, T., Abedi, H., \& Aguilar, L.\ 2015, \mnras, 447, 218
\bibitem[Saffer et al.(1994)]{Saffer94} Saffer, R.~A., Bergeron, P., Koester, D., \& Liebert, J. 1994, ApJ, 432, 351
\bibitem[Sales Silva et al.(2016)]{sal16} Sales Silva, J.V., Carraro, G., Anthony-Twarog,B.J., Moni Bidin, C., Costa, E., \& Twarog, B.A. 2016, AJ, 151, 6
\bibitem[Santrich et al.(2013)]{san13} Santrich, O.~J.~K., Pereira, C.~B., \& Drake, N.~A.\ 2013, \aap, 554, A2
\bibitem[Sbordone et al.(2005)]{sbo05} Sbordone, L., Bonifacio, P., Marconi, G., Zaggia, S., \& Buonanno, R. 2005, A\&A, 430, L13
\bibitem[Sbordone et al.(2007)]{sbo07} Sbordone, L., Bonifacio, P., Buonanno, R., Marconi, G., Monaco, L., \& Zaggia, S. 2007, A\&A, 465, 815
\bibitem[Sbordone et al.(2015)]{sbo15} Sbordone, L., Monaco, L., Moni Bidin, C., et al. 2015, A\&A, 579, 104
\bibitem[Schlegel et al.(1998)]{Schlegel98} Schlegel, D.~J., Finkbeiner, D.~P., \& Davis, M. 1998, ApJ, 500, 525
\bibitem[Sneden(1973)]{sne73} Sneden, C. 1973, ApJ, 184, 839
\bibitem[Sneden et al.(2004)]{Sneden04} Sneden, C., Ivans, I.~I., \& Fulbright, J.~P. 2004, in Origin and Evolution of the Elements, 17
\bibitem[Tonry \& Davis (1979)]{ton79} Tonry, J., \& Davis, M. 1979, AJ, 84, 1511
\bibitem[Turner(1976)]{turner76} Turner, D.~G., 1976, \aj, 81, 97
\bibitem[Turner(1979)]{turner79} Turner, D.G., 1979, \pasp, 91, 642
\bibitem[Turner(1983)]{turner83} Turner, D.~G., 1983, JRASC, 77, 31
\bibitem[Turner et al.(2011)]{turner11} Turner, D.~G., MacLellan, R.~F., Henden, A.~A., \& Bernikov, L.~N. 2011, RMxAA, 47, 345
\bibitem[Turner et al.(2014)]{turner14} Turner, D.~G., Majaess, D.~J., \& Balam, D.~D. 2011, CJP, 92, 1696
\bibitem[van Dokkum(2001)]{dok01} van Dokkum, P.~G. 2001, PASP, 113, 1420
\bibitem[Vazquez et al.(2008)]{vazquez08} Vazquez, R.A., May, J., Carraro, G., Bronfman, L., Moitinho, A., Baume, G., 2008, ApJ, 672, 930
\bibitem[Wan et al.(2015)]{wan15} Wan, J.-C., Liu, C., Deng, L.-C., et al.\ 2015, Research in Astronomy and Astrophysics, 15, 1166
\bibitem[Wiese et al.(1969)]{wie69} Wiese, W.~L., Smith, M.~W., \& Miles, B.~M.\ 1969, NSRDS-NBS, Washington, D.C.: US Department of Commerce, National Bureau of Standards, 1969
\bibitem[Woosley \& Weaver(1995)]{woo95} Woosley, S.~E., \& Weaver, T.~A. 1995, ApJS, 101, 181
\bibitem[Xu et al.(2015)]{xu15} Xu, Y., Newberg, H.~J., Carlin, J.~L., et al.\ 2015, \apj, 801, 105
\bibitem[Yong et al.(2005)]{yon05} Yong, D., Carney, B.~W., Teixera de Almeida, M.~L., \& Pohl, B.~L. 2006, AJ, 131, 2256
\bibitem[Yong et al.(2006)]{yon06} Yong, D., Carney, B.~W., Teixera de Almeida, M.~L., \& Pohl, B.~L.\ 2006, AJ, 131, 2256
\end{thebibliography}
\end{document}